\documentclass[12pt,english]{article}

\usepackage[T1]{fontenc}
\usepackage[utf8]{inputenc}
\usepackage{libertine}
\usepackage{inconsolata}

\usepackage{amsfonts}
\usepackage{amsmath,amssymb,amsthm}

\usepackage{hyperref}
\hypersetup{
  citecolor=blue,
  colorlinks=true,
}

\usepackage{tikz}

\usetikzlibrary{arrows,automata}
\usetikzlibrary{positioning}

\usepackage{mathtools}
\usepackage{braket}
\usepackage{stmaryrd}
\usepackage{mathdots}
\usepackage[export]{adjustbox}
\usepackage{booktabs}
\usepackage{caption}
\usepackage{subcaption}
\usepackage{color} 
\usepackage{algorithm}
\usepackage{fullpage}
\usepackage{graphicx}
\usepackage{algorithmic}
\ifpdf
  \DeclareGraphicsExtensions{.eps,.pdf,.png,.jpg}
\else
  \DeclareGraphicsExtensions{.eps}
\fi

\usepackage[english]{babel}
\usepackage{authblk}

\makeatletter
\newenvironment{proofof}[1]{\par
  \pushQED{\qed}%
  \normalfont \topsep6\p@\@plus6\p@\relax
  \trivlist
  \item[\hskip\labelsep
        \bfseries
    Proof of #1\@addpunct{.}]\ignorespaces
}{%
  \popQED\endtrivlist\@endpefalse
}
\makeatother

\title{Quantum circuits synthesis using Householder transformations}

\date{}

\author[1,3]{Timothée Goubault de Brugière}
\author[1]{Marc Baboulin}
\author[2]{Benoît Valiron}
\author[3]{Cyril~Allouche}

\affil[1]{Université Paris-Saclay, CNRS,                  Laboratoire de recherche en informatique, 91405, Orsay, France}
\affil[2]{Université Paris-Saclay, CNRS, CentraleSupélec, Laboratoire de Recherche en Informatique, 91405, Orsay, France}
\affil[3]{Atos Quantum Lab, Les Clayes-sous-Bois, France}

\begin{document}

\maketitle

\begin{abstract}
The synthesis of a quantum circuit consists in decomposing a unitary matrix into a series of elementary operations. In this paper, we propose a circuit synthesis method based on the QR factorization via Householder transformations. 
We provide a two-step algorithm: during the first step we exploit 
  the specific structure of a quantum operator to compute its QR factorization,
  then the factorized matrix is used to produce a quantum circuit. We analyze several costs (circuit size and computational time) and compare them to existing techniques from the literature. For a final quantum circuit twice as large as the one obtained by the best existing method, we accelerate the computation by orders of magnitude.
\end{abstract}

\section{Introduction}

In the 1980s the notion of a quantum computer emerged as a response to
the announced limitation of conventional computers in terms of
computing power. Feynman~\cite{feynman1982simulating}, then Deutsch~\cite{Deutsch97} announced and theorized the
first foundations of this new paradigm that must override our current
machines. Ten years later, we
saw the first concrete algorithms capable of achieving this quantum
supremacy: the Grover algorithm theoretically enables us to search into an
unstructured database quadratically faster than in the conventional
case~\cite{Grover:1996:FQM:237814.237866} and the Shor algorithm
is expected to be able to break RSA, jeopardizing the
security of current encryption tools~\cite{chen2016report,doi:10.1137/S0036144598347011}.
Quantum computing is now a research topic of growing interest and many algorithms
are designed in numerous fields to try to surpass classical
computers. Examples are various: machine learning
\cite{biamonte2017quantum,kerenidis2017recommendation},  
linear algebra \cite{kerenidis2017quantum,larose2019variational,peruzzo2014variational}, backtracking algorithms \cite{montanaro2015quantum} or even combinatorial optimization \cite{farhi2014quantum}. The interaction between classical computing and 
quantum computing is also studied, leading to hybrid quantum-classical computers \cite{suchara2018hybrid}.
Behind all these new algorithms lies a common formalism: the quantum circuit.
Developed by Yao~\cite{yao1993quantum}, the concept of a quantum circuit remains so far
the preferred way to describe quantum algorithms. Similarly to the
compilation in classical computing, transforming a high level concept
--- or more generally a concept unknown to the hardware --- into a
sequence of basic instructions understandable by the machine is a central
problem. In quantum computing, everything can be modeled with notions
of linear algebra: states are vectors, operators are unitary
matrices. The compilation problem can be formalized as the
transformation of a unitary matrix into a quantum circuit consisting
of elementary (unitary) operations admissible by the hardware and
referred to as elementary quantum gates. The development of quantum
algorithmics has fostered the emergence of high-level
languages~\cite{ScaffCC,Liquid,Quipper} to efficiently describe and
program concrete instances of quantum algorithms. With the limited
resources that are going to be available at first for quantum
computers, it is crucial to design an automated compilation process
minimizing the classical and quantum resources used by a given quantum
program. 

When turning a unitary matrix into a quantum circuit, several aspects
have to be considered. First, one has to decide on the set of
admissible elementary operations. Then, one has to choose the
resources to be minimized: are we interested in the smallest possible
circuit, or are we also considering the classical resources used to
produce the circuit and the time required to do so? The former
problem is very theoretical and math-oriented.
An operator acting on $n$ qubits is represented by a matrix of size
$2^n\times2^n$. Generating a circuit from an arbitrary matrix is
therefore a problem that scales exponentially in $n$ in general, and the problem of
finding the smallest possible circuit for a particular operator
remains challenging~\cite{knill1995approximation}.
Nonetheless, several techniques have been developed to this end using
e.g., decomposition methods~\cite{PhysRevA.52.3457,
  doi:10.1109/5992.908999, mottonen2005decompositions,
  PhysRevLett.73.58, 1629135}. The resulting number of gates however still
lies within a factor of 2 of the theoretical lower
bound~\cite{Bullock:2004:AOC:2011572.2011575}.
We are currently in the NISQ (Noisy Intermediate-Scale Quantum) era \cite{NISQC}: the quantum hardware is noisy and it is hard to perform long computations. In this paper we foresee the future of the NISQ era where full fault tolerant quantum computation will be available. With the advent of such systems, we believe that the synthesis of generic operators on small to medium register size will become critical. For example one can already get a glimpse of such issue in quantum machine learning problems \cite{biamonte2017quantum}. Meanwhile, we can also rely on post processing methods that integrate the presence of noise in the hardware and make the connection between ideal quantum circuit synthesis and the hardware constraints \cite{shi2019optimized}.

Instead of only focusing on the size of the circuit, one can consider
the problem in its globality and also take into account the quantity
of classical resources needed, and in particular the time it takes to
generate the circuit. 
Such optimization is particularly useful, e.g. when one has to compile a continuous stream of quantum circuits on the fly or when the quantum operator is parameterized and one has to recompile the parameters of the resulting quantum circuit every time the operator changes. Improving the compilation time also allows to reach larger problem sizes.
This aspect of optimization is a recent
topic of research~\cite{amy2013meet, 2058-9565-4-1-015004, 2058-9565-1-1-015003, Continuous} and is
the focus of our paper. 

\subsection{Contributions}
The main contributions of the paper are as follows.
\begin{itemize}
\item We adapt the well known and numerically stable QR factorization
  based on Householder transformations~\cite{GVL96} to the
  factorization of unitary matrices. The adaptation heavily relies on the
  specific structure of unitary matrices. We exhibit a significant
  theoretical and practical speedup of our specific QR
  algorithm compared to the unmodified QR routine and the usual
  technique for quantum circuit synthesis based on the quantum Shannon
  decomposition (QSD)~\cite{1629135}.
\item We propose a complete circuit synthesis method using this
  specialized QR decomposition with a complexity analysis for circuit size and arithmetical operations.
  If some existing theoretical and
  experimental works for quantum circuits synthesis with Householder transformations 
  have been undertaken~\cite{doi:10.1063/1.3466798, PhysRevA.74.022323,
    doi:10.1139/cjp-2015-0490}, to our knowledge
  none has proposed an implementation method and a final circuit
  construction with clearly defined properties. Overall, our technique is
  faster than the QSD-based method while providing
  circuits twice as large\footnote{This extra cost in the
    final quantum circuit is not negligible, especially when
    considering the current limitations of the quantum
    hardware. It may
    be possible that the gain in the classical process will not
    compensate the execution time of the twice as large quantum
    circuit on real hardware. However, we can handle problem sizes that were
    unreachable before with the QSD, regardless of the quality of the hardware. We believe our approach highlights the tradeoffs between two measures of complexity (circuit size/compilation time) and that this has to be taken in consideration when synthesizing generic quantum circuits.}.
\item We backup our approach with benchmarks on multicore and GPU architectures for
 random unitary matrices operating on up to 15 qubits. 
\end{itemize}

\noindent
This is a preprint submitted for publication.

\subsection{Plan of the paper}
The plan of this paper is as follows. In Section~\ref{sec:background}
we give some background about quantum computing, quantum circuits and
the issues in quantum compilation. Then we detail the new adapted
Householder algorithm in Section~\ref{sec:householder} and we explain
in Section~\ref{sec:circuit} how to convert this factorization into a
quantum circuit. Section~\ref{sec:exp} presents the performance obtained on multicore and GPU architectures by our algorithm. We also compare our results with 
a reference algorithm based on the Quantum Shannon Decomposition
method. We conclude in Section~\ref{sec:concl}.

\subsection{Notations}
Throughout this paper we will use the following notations. $\mathcal{U}(k)$ denotes the set of 
unitary matrices of size $k$, i.e. 
$\mathcal{U}(k) = \{ M \in  \mathbb{C}^{k \times k} \; | \; M^{\dag}M = I \}$,
where $I$ is the identity matrix and $M^{\dag}$ is the conjugate transpose of the matrix $M$. 
The notation $\|\cdot\|$ refers to the Euclidean norm of a vector and $e_i$ is the $i$th canonical vector.
The term {\it flops} stands for {\it floating-point operations} and the flop count evaluates the volume of work in a computation. Unless otherwise specified theses flops are given in complex arithmetic. The linear algebra formulas will be presented using matlab-like notations.

\section{Background}\label{sec:background}

The core of quantum computation consists in encoding information on
the state $\ket\phi$ of a quantum system. The computational model is
derived from the laws of quantum mechanics: the state $\ket\phi$ is
represented by a normalized column vector in a (finite dimensional)
Hilbert space $\mathbb{C}^k$.
The allowed transformations one can perform on $\ket\phi$ can be
derived from the Schrödinger equation. In this paper we
focus on \emph{unitary} transformations. A quantum operator
acting on the vector $\ket\phi\in\mathbb{C}^k$ is therefore in this
paper regarded as a unitary matrix $U \in \mathcal{U}(k)$. After
computation, the resulting state is $U\ket\phi$. A \emph{sequential}
application of two transformations $U$ and $V$ yields the state
$V(U\ket\phi) = (VU)\ket\phi$ and corresponds to a matrix multiplication.

The basic unit of information in quantum computation is the quantum
bit, also called qubit. It is encoded by a two-level quantum system
(e.g., the spin of an electron) whose state can be in a linear
superposition of both levels --- called the basis states --- according to the
laws of quantum mechanics. We usually write $\ket{0}$ to represent the
first basis state and $\ket{1}$ the second one (to follow the analogy
with the classical case). The general form $\ket{\psi}$ of the state of a qubit is
then the linear combination of these basis elements $\ket0$ and $\ket1$ (also
called ``superposition''):
\[ \ket{\psi} = \alpha \ket{0} + \beta \ket{1}\]
where $\alpha, \beta$ are complex numbers such that $|\alpha|^2 +
|\beta|^2 = 1$. In other words, the state of a qubit is mathematically equivalent
to a unit vector $(\begin{smallmatrix} \alpha \\
  \beta \end{smallmatrix}) \in \mathbb{C}^2$ and the basis states are the
usual basis vectors 
\[ \ket{0} = \begin{pmatrix} 1 \\ 0 \end{pmatrix} \; \; \; \ket{1}
  = \begin{pmatrix} 0 \\ 1 \end{pmatrix}. \] The qubit is not the only
logical unit possible in quantum computing: using 3-level systems (by
adding the basic state $\ket{2}$) one can manipulate qutrits\,; more
generally with a $d$-level system we talk about qudits. However the
research in quantum computing today uses mostly qubits.

The state of the quantum system consisting of the \emph{combination}
of two systems $A$ and $B$ resides in the Kronecker (tensor) product 
of the space of states of $A$ and the space of states of $B$. In
particular,
to encode $n$ qubits, one can use $n$ two-level systems that together
can be seen as a single $2^n$ level system. The
evolution of this system is governed by the left multiplication by
unitary matrices in $\mathcal{U}(2^n)$. The basis vectors of the space
$\mathbb{C}^{2^n}$ are of the form 
$\ket{x_1}\otimes\cdots\otimes\ket{x_n}$ with  $x_i=0$ or $1$.
The usual ordering of the basis states corresponds to the
lexicographic order. For example, in the case of two qubits the basis
states are
\[
  \ket0\otimes\ket0 =
  \left(\begin{smallmatrix}1\\0\\0\\0\end{smallmatrix}\right),~
  \ket0\otimes\ket1 =
  \left(\begin{smallmatrix}0\\1\\0\\0\end{smallmatrix}\right),~
  \ket1\otimes\ket0 =
  \left(\begin{smallmatrix}0\\0\\1\\0\end{smallmatrix}\right),~
  \ket1\otimes\ket1 =
  \left(\begin{smallmatrix}0\\0\\0\\1\end{smallmatrix}\right).
\]
To combine operators acting
on distinct subsystems, we again use the tensor
product.
If $\ket{\psi}$ (resp. $\ket{\phi}$) is an $n$-qubit (resp. $m$-qubit) state
and one applies an operator $A$ on
$\ket{\psi}$ (resp. $B$ on $\ket\phi$)
then using the global system on $n+m$ qubits it is equivalent to applying
the operator $A \otimes B$ on the state
$\ket{\psi} \otimes \ket{\phi}$, where $\otimes$ denotes the Kronecker
product~\cite{LOAN200085}.

When a state on $n$ qubits cannot be written as a tensor product of two substates then the state is said to be entangled. The Bell states are simple examples of entangled states on two qubits, one of them is defined by 
\[ \ket{\Phi} = \frac{1}{\sqrt{2}} \left( \ket{0}\otimes\ket{0} + \ket1\otimes\ket1 \right) \] 
and one can check that it cannot be expressed as the tensor product of two one-qubit states. Entanglement is believed to be a key in the quantum supremacy over classical computation~\cite{Jozsa2011} and research is performed to better understand its role, for example by giving a measure of how entangled a state is \cite{PhysRevLett.91.147902}. An operator that can produce entangled states is said to be an \emph{entangling} operator.

Beside composition and combination, a third operation is usually
considered: an operation can be \emph{controlled}. If
$M\in\mathcal{U}(2^n)$ is an operator acting on $n$ qubits, there are
two canonical operations on $n+1$ qubits: the positively-controlled-$M$
defined as the block matrix
$(\begin{smallmatrix}I&0\\0&M\end{smallmatrix})$ and the
negatively-controlled-$M$, defined as
$(\begin{smallmatrix}M&0\\0&I\end{smallmatrix})$. Both block-matrices
are operators in $\mathcal{U}(2^{n+1})$. The former sends
$\ket0\otimes\ket\phi$ to $\ket0\otimes\ket\phi$ and
$\ket1\otimes\ket\phi$ to $\ket1\otimes(M\ket\phi)$. The latter does
the opposite: it sends $\ket0\otimes\ket\phi$ to $\ket0\otimes(M\ket\phi)$ and
$\ket1\otimes\ket\phi$ to $\ket1\otimes\ket\phi$. 

An important notion is the {\it preparation} and {\it de-preparation} of states. Preparing a state $\ket{\Phi}$ consists in applying an operator $U$ to the state $\ket0$ to obtain the state $\ket{\Phi}$. Conversely, the de-preparation of the state $\ket{\Phi}$ consists in applying $U^{\dag}=U^{-1}$ to obtain the state $\ket0$.

\medskip
\noindent
\paragraph{Quantum gates}
Though the theory allows arbitrary unitary matrices, the physical hardware
is usually only capable of handling a
fixed set of unitary matrices operating on one or two qubits. These elementary
matrices are called \emph{quantum gates}, and we can mention the
following (see Table~\ref{tab:gates}):
\begin{itemize}
	\item the Pauli operators $X,Y,Z$ (the $X$ gate is equivalent to the classical NOT gate),
	\item the Hadamard gate $H$ which enables us to transform a pure state (i.e. $\ket{0}$ or $\ket{1}$) into an equal superposition of $\ket{0}$ and $\ket{1}$,
	\item the continuous set of elementary rotations $R_x, R_y, R_z$ defined by 
          \[
            R_{G}(\alpha) = \cos(\alpha/2) I_2 - i \;
            \sin(\alpha/2) G \text{ with } G \in \{X,Y,Z\}
          \]
          where $X,Y,Z$ are the Pauli operators and $i$ is the unit imaginary number.
   \item the continuous set of phase gates defined by 
   \[ Ph(\theta) = \begin{pmatrix} 1 & 0 \\ 0 & e^{i\theta} \end{pmatrix} \]
   adding a phase to the state $\ket{1}$ ; among this set two gates
   are of particular use: the gate $T$ ($\theta = \pi/4$) and the
   gate $S$ ($\theta = \pi/2$).
   Note that $Ph(\theta)$ is simply $R_z(\theta)$ modulo a global
   phase $e^{-i\frac\theta2}$.
\end{itemize}~\\
Amongst the frequently used 2-qubit gates, one can name the CNOT-gate,
which is the positively-controlled X-gate, and the SWAP gate, flipping
the state of two qubits. Other examples of commonly encountered gates
are controlled-rotations with arbitrary angles.
\begin{table}[tb]
  \[
    \begin{array}{c}
      \begin{array}{c}
        \begin{pmatrix} 0 & 1 \\ 1 & 0 \end{pmatrix}
        \\[2.3ex]
        X
      \end{array}
      \quad
      \begin{array}{c}
        \begin{pmatrix} 0 & -i \\ i & 0 \end{pmatrix}
        \\[2.3ex]
        Y
      \end{array}
      \quad
      \begin{array}{c}
        \begin{pmatrix} 1 & 0 \\ 0 & -1 \end{pmatrix}
        \\[2.3ex]
        Z
      \end{array}
      \\ \\
      \begin{array}{c}
        \frac{1}{\sqrt{2}}\begin{pmatrix} 1 & 1 \\ 1 & -1 \end{pmatrix}
        \\[2.3ex]
        H
      \end{array}
      \quad
      \begin{array}{c}
        \begin{pmatrix} 1 & 0 \\ 0 & i \end{pmatrix}
        \\[2.3ex]
        S
      \end{array}
      \quad
      \begin{array}{c}
        \begin{pmatrix} 1 & 0 \\ 0 & e^{i\pi/4} \end{pmatrix}
        \\[2.3ex]
        T
      \end{array}
    \end{array}
    ~
    \begin{array}{c}
      \begin{pmatrix}
        1 & 0 & 0 & 0 \\
        0 & 1 & 0 & 0 \\
        0 & 0 & 0 & 1 \\
        0 & 0 & 1 & 0
      \end{pmatrix}
      \\[5ex]
      \text{CNOT}
    \end{array}
    \quad
    \begin{array}{@{}c@{}}
      \begin{pmatrix}
        1 & 0 & 0 & 0 \\
        0 & 0 & 1 & 0 \\
        0 & 1 & 0 & 0 \\
        0 & 0 & 0 & 1
      \end{pmatrix}
      \\[5ex]
      \text{SWAP}
    \end{array}    
  \]
  \caption{Usual elementary unitary matrices for representing quantum gates.}
  \label{tab:gates}
\end{table}

\medskip
\noindent
\paragraph{Quantum circuit}
The usual graphical language for representing composition and combination
of operator is the 
equivalent of the boolean circuit for classical computing: the quantum
circuit. A quantum circuit consists in a series of parallel,
horizontal wires on which are attached boxes. Each wire corresponds to
a qubit, and vertical combination corresponds to the Kronecker
(tensor) product. The circuit is read from left to
right and each box corresponds to a quantum gate (i.e., a unitary
operator)
applied on the corresponding 
qubits.
Controlled-gates have a special representation: the controlling qubit
is represented with a bullet if the control is positive and a circle
if the control is negative. A vertical line then connects the
controlling qubit to the gate to be controlled. The notation
easily extends to multiple controls.

\begin{figure}[t]
\newcommand{\xs}{1.4}
\newcommand{\gate}[4]{\draw[fill=white] (\xs*#1-#3,#2-#3) rectangle (\xs*#1+#3,#2+#3); \node at (\xs*#1,#2) {#4} ;}%
\newcommand{\cgate}[5]{\draw[fill=black] (\xs*#1,#3) circle (0.1) ; \draw (\xs*#1,#2) -- (\xs*#1,#3) ; \draw[fill=white] (\xs*#1-#4,#2-#4) rectangle (\xs*#1+#4,#2+#4); \node at (\xs*#1,#2) {#5} ;}%
\newcommand{\cgatex}[6]{\draw[fill=black] (\xs*#1,#3) circle (0.1) ; \draw (\xs*#1,#2) -- (\xs*#1,#3) ; \draw[fill=white] (\xs*#1-#6,#2-#4) rectangle (\xs*#1+#6,#2+#4); \node at (\xs*#1,#2) {#5} ;}%
\newcommand{\qline}[3]{\draw (0,#1) node [anchor=east] {#3} -- (\xs*#2,#1) ;}
\centering
\begin{tikzpicture}[x=4ex,y=-4ex]
 \qline{1}{11}{$q_0$}\qline{2}{11}{$q_1$}\qline{3}{11}{$q_2$}\qline{4}{11}{$q_3$}
 \gate{1}{1}{0.4}{$H$}
 \cgate{2}{1}{2}{0.4}{$S$}
 \cgate{3}{1}{3}{0.4}{$T$}
 \cgatex{4}{1}{4}{0.4}{\scalebox{.8}{$\mathit{Ph}(\frac\pi8)$}}{.65}
 \gate{5}{2}{0.4}{$H$}
 \cgate{6}{2}{3}{0.4}{$S$}
 \cgate{7}{2}{4}{0.4}{$T$}
 \gate{8}{3}{0.4}{$H$}
 \cgate{9}{3}{4}{0.4}{$S$}
 \gate{10}{4}{0.4}{$H$}
\end{tikzpicture}
\caption{Quantum circuit for the Quantum Fourier Transform}
\label{qft}
\end{figure}
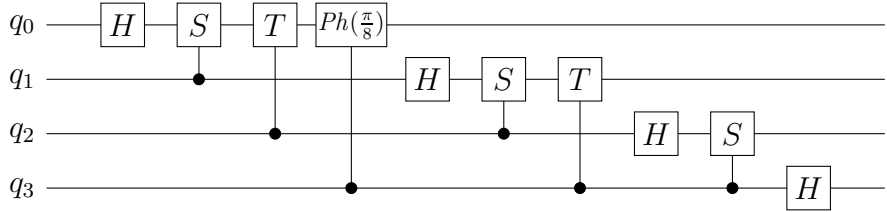

As an example of quantum circuit combining
several gates together, the so-called
Quantum Fourier Transform~\cite{nielsen2011quantum}
is represented in Figure~\ref{qft}. It
enables us to visualize the use of elementary gates: $H$, $S$, $T$,
phase-gate, and positive controls.

\medskip
\noindent
\paragraph{Universality}
We say that a set of gates is universal if any quantum operator,
acting on any number of qubits, can be implemented as a sequence of
gates from this set (see e.g. \cite[Sec. 4.5]{nielsen2011quantum} for
a complete discussion on the matter). A fundamental (theoretical) result claims that it is possible to
realize any operator only with the set of one-qubit gates and one
``sufficiently entangling'' 2-qubit gate such as the
CNOT~\cite{PhysRevLett.89.247902}. 
To be able to implement any quantum algorithm with a given piece of
hardware, it is therefore necessary to first find a universal set of
technologically implementable gates. For instance the IBM quantum
machine with superconducting qubits uses only special unitary gates on one qubit and the CNOT
\cite{PhysRevA.94.012314}. A technology using trapped ions will have
other gates available like the MS gate \cite{1367-2630-19-2-023035, 1367-2630-15-12-123012}. 
Linear quantum optics will instead focus on
CNOTs and one-qubit gates~\cite{kok2007linear}.

But having an implementable universal set of gates is not enough: if we are given a quantum operator
as a unitary matrix, one
also has to find a way to turn the desired unitary matrix acting on a
potentially large number of qubits into a quantum circuit made of
local, elementary gates.
This problem is known as \emph{circuit synthesis}, or equivalently,
\emph{compilation} of the unitary matrix into a circuit. 
Note that the state preparation is a special case of circuit synthesis where we want to synthesize only the first column of a unitary matrix.

In this paper, we focus on
the set consisting of the CNOT gate and all the one-qubit gates. The
continuous aspect of the set makes it amenable to linear algebra
operations, and yet it can easily be mapped to other universal gate\textcolor{red}{-}sets \cite{barenco1995universal, deutsch1995universality,
  PhysRevA.51.1015, PhysRevLett.75.346}, using for example the
Solovay-Kitaev theorem \cite{Dawson:2006:SA:2011679.2011685} or more
recent techniques~\cite{kliuchnikov2013fast, ross2016optimal, selinger2012efficient}.

\begin{figure}[t]
\centering
\newcommand{\xs}{1}\newcommand{\gs}{.4}\newcommand{\bs}{.11}
\newcommand{\ccgate}[4]{
  \draw (#1,0) -- (#1,2) ; 
  \draw[fill=#2] (#1,0) circle (\bs) ; 
  \draw[fill=#3] (#1,1) circle (\bs) ;
  \draw[fill=white] (#1-\gs,2-\gs) rectangle (#1+\gs,2+\gs); 
  \node at (#1,2) {\scalebox{.8}{#4}} ;}%
\newcommand{\bbgate}[2]{
  \draw (#1,0) -- (#1,2) ; 
  \draw[fill=white] (#1-\bs,-\bs) rectangle (#1+\bs,0+\bs) ; 
  \draw[fill=white] (#1-\bs,1-\bs) rectangle (#1+\bs,1+\bs) ;
  \draw[fill=white] (#1-\gs,2-\gs) rectangle (#1+\gs,2+\gs); 
  \node at (#1,2) {\scalebox{.8}{#2}} ;}%
\begin{minipage}{1.8cm}
\begin{tikzpicture}[x=4ex,y=-4ex]
\node at (0.2,2) {$/$} ;
\draw (0,0) -- (2,0) ;
\draw (0,1) -- (2,1) ;
\draw (0,2) -- (2,2) ;
\bbgate{1}{$A$}
\end{tikzpicture}
\end{minipage}
\qquad
$\equiv$
\qquad
\begin{minipage}{4cm}
\begin{tikzpicture}[x=4ex,y=-4ex]
\node at (0.2,2) {$/$} ;
\draw (0,0) -- (5,0) ;
\draw (0,1) -- (5,1) ;
\draw (0,2) -- (5,2) ;
\ccgate{1}{white}{white}{$A_0$}
\ccgate{2}{white}{black}{$A_1$}
\ccgate{3}{black}{white}{$A_2$}
\ccgate{4}{black}{black}{$A_3$}
\end{tikzpicture}
\end{minipage}
\caption{Circuit equivalence for a multiplexor}
\label{mult}
\end{figure}
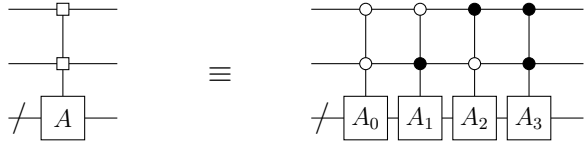

\medskip
\noindent
\paragraph{Multiplexors}
Some circuit structures are expressive enough to be reused in other
constructions, thus helping during the compilation process. The most
important of them is the multiplexor \cite{PhysRevA.71.052330, PhysRevLett.93.130502, 1629135}.
It can be regarded as a generalization
of controlled gates because it applies a different operator for each
value of the control qubits. For instance a multiplexor controlled by
two qubits and acting on $k$ qubits has the matrix block structure
\[
 A = \left(\begin{matrix}
      A_0&0&0&0\\
      0& A_1 &0&0\\
      0&0& A_2&0\\
      0&0&0& A_3
    \end{matrix}\right)
\]
where $A_0$ is a $k$-qubit operator that is applied if both control
qubits are 0, $A_1$ is applied if the first control qubit is 0 and the
second one is 1, \emph{etc}. The graphical representation of a
multiplexor is illustrated in Figure~\ref{mult} with the
correspondence between $A$ and a succession of multi-controlled gates.
In the circuit, the crossed-out line stands for a several
qubits (in this case, $k$ qubits).

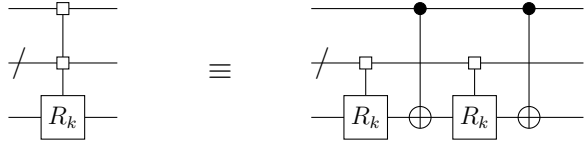
\begin{figure}[t]
  \centering
\newcommand{\xs}{1}\newcommand{\gs}{.4}\newcommand{\bs}{.11}\newcommand{\nots}{.2}
\newcommand{\bbgate}[2]{
  \draw (#1,0) -- (#1,2) ; 
  \draw[fill=white] (#1-\bs,-\bs) rectangle (#1+\bs,0+\bs) ; 
  \draw[fill=white] (#1-\bs,1-\bs) rectangle (#1+\bs,1+\bs) ;
  \draw[fill=white] (#1-\gs,2-\gs) rectangle (#1+\gs,2+\gs); 
  \node at (#1,2) {\scalebox{.8}{#2}} ;}%
\newcommand{\bgate}[2]{
  \draw (#1,1) -- (#1,2) ; 
  \draw[fill=white] (#1-\bs,1-\bs) rectangle (#1+\bs,1+\bs) ; 
  \draw[fill=white] (#1-\gs,2-\gs) rectangle (#1+\gs,2+\gs); 
  \node at (#1,2) {\scalebox{.8}{#2}} ;}%
\newcommand{\cnot}[3]{
  \draw (#1,#3) circle (\nots) ;
  \draw (#1,#2) -- (#1,#3+\nots) ;
  \draw[fill=black] (#1,#2) circle (\bs) ;
}
\begin{minipage}{1.8cm}
\begin{tikzpicture}[x=4ex,y=-4ex]
\node at (0.2,1) {$/$} ;
\draw (0,0) -- (2,0) ;
\draw (0,1) -- (2,1) ;
\draw (0,2) -- (2,2) ;
\bbgate{1}{$R_k$}
\end{tikzpicture}
\end{minipage}
\qquad
$\equiv$
\qquad
\begin{minipage}{4cm}
\begin{tikzpicture}[x=4ex,y=-4ex]
\node at (0.2,1) {$/$} ;
\draw (0,0) -- (5,0) ;
\draw (0,1) -- (5,1) ;
\draw (0,2) -- (5,2) ;
\bgate{1}{$R_k$}
\cnot{2}{0}{2}
\bgate{3}{$R_k$}
\cnot{4}{0}{2}
\end{tikzpicture}
\end{minipage}
\caption{Decomposition of a rotation multiplexor}
\label{mult_ry}
\end{figure}

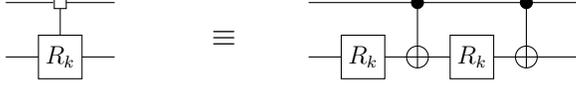
\begin{figure}[t]
\centering
\newcommand{\xs}{1}\newcommand{\gs}{.4}\newcommand{\bs}{.11}\newcommand{\nots}{.2}
\newcommand{\bgate}[2]{
  \draw (#1,1) -- (#1,2) ; 
  \draw[fill=white] (#1-\bs,1-\bs) rectangle (#1+\bs,1+\bs) ; 
  \draw[fill=white] (#1-\gs,2-\gs) rectangle (#1+\gs,2+\gs); 
  \node at (#1,2) {\scalebox{.8}{#2}} ;}%
\newcommand{\gate}[2]{
  \draw[fill=white] (#1-\gs,2-\gs) rectangle (#1+\gs,2+\gs); 
  \node at (#1,2) {\scalebox{.8}{#2}} ;}%
\newcommand{\cnot}[3]{
  \draw (#1,#3) circle (\nots) ;
  \draw (#1,#2) -- (#1,#3+\nots) ;
  \draw[fill=black] (#1,#2) circle (\bs) ;
}
\begin{minipage}{1.8cm}
\begin{tikzpicture}[x=4ex,y=-4ex]
\draw (0,1) -- (2,1) ;
\draw (0,2) -- (2,2) ;
\bgate{1}{$R_k$}
\end{tikzpicture}
\end{minipage}
\qquad
$\equiv$
\qquad
\begin{minipage}{4cm}
\begin{tikzpicture}[x=4ex,y=-4ex]
\draw (0,1) -- (5,1) ;
\draw (0,2) -- (5,2) ;
\gate{1}{$R_k$}
\cnot{2}{1}{2}
\gate{3}{$R_k$}
\cnot{4}{1}{2}
\end{tikzpicture}
\end{minipage}
\caption{Decomposition of a rotation multiplexor with one control qubit}
\label{mult_ry_bis}
\end{figure}

The problem of decomposing a multiplexor into elementary gates admits
algorithms with varying numerical costs depending on the choice of
elementary gates~\cite{PhysRevLett.93.130502, 1629135}. In the case of
a multiplexor applying only one kind of elementary rotations
(e.g. along one of the axis $X$, $Y$ or $Z$ -- we call such a structure a rotation
multiplexor) the transition from the angles of the multiplexors to the
angles in the corresponding quantum circuits can be performed via a single matrix-vector
product \cite{PhysRevLett.93.130502}. Moreover, the decomposition is
much simpler than the general case shown in Figure~\ref{mult}:
the decomposition of an $R_k$-multiplexor controlled by $n$ qubits into two multiplexors controlled by $n-1$ qubits has the shape shown in Figure~\ref{mult_ry} and the special case with one control qubit is shown in Figure~\ref{mult_ry_bis} (where we omit
angles for legibility). Such decompositions can be applied recursively and by removing some CNOT gates that cancel (see Figure 2 in \cite{1629135} for more details)
we obtain a final quantum circuit composed of $2^{n-1}$ CNOTs and $2^{n-1}$ elementary rotations. For multiplexors in $SU(2)$ the decomposition given in Figure~\ref{mult_ry} remains valid up to a diagonal matrix that replaces the last CNOT gate. Hence, without considering this extra gate --- for our purpose we will be able to remove it --- we need $2^{n-1}-1$ CNOTs and $2^{n-1}$ generic one-qubit gates to implement an $SU(2)$-multiplexor.

\medskip
\noindent
\paragraph{Quantum state preparation}
A common method for preparing a generic quantum state on $n$ qubits
consists in applying a series of operations such that we are left with
the preparation of a quantum state on $n-1$ qubits, and we repeat the
process until we have to prepare only a one-qubit state. Desentangling
the first qubit is for instance equivalent to zeroing the second half of the components of the corresponding vector $\Psi$. To do so, one can apply for each bitstring $s \in F_2^{n-1}$ a specific two-qubit operation $U_s$  on the first qubit such that $U_s(\Psi_{0s} \ket{0s} + \Psi_{1s} \ket{1s}) = \Psi'_{0s} \ket{0s}$. Then the global operator $\bigoplus_s U_s$ can be implemented either by applying successively one $R_z$-multiplexor and one $R_y$-multiplexor, both on the first qubit and controlled by the $n-1$ other ones, or by applying one $SU(2)$-multiplexor, still on the first qubit and controlled by the other qubits \cite{mottonen2005decompositions,1629135}. In the case where we only use $R_y$ and $R_z$ multiplexors, we can simply repeat the operation on the $n-1$ remaining qubits. Overall we need to implement two rotation multiplexors on $n$ qubits, two on $n-1$ qubits etc. for a total of $2 \times \sum_{k=2}^n 2^{k-1} \approx 2^{n+1}$ CNOTs and $2 + 2 \times \sum_{k=2}^n 2^{k-1} \approx 2^{n+1}$ elementary rotations. Some optimizations can decrease the CNOT-count by a linear term in $n$ but we focus on the asymptotic complexity. When using multiplexors in $SU(2)$, we remark that the additional diagonal gate in the synthesis of the multiplexor can be merged with the remaining quantum state as adding phases to each component of the state will not change the number of nonzero elements. So preparing a quantum state with multiplexors in $SU(2)$ requires to implement one $SU(2)$-multiplexor on $n$ qubits, one on $n-1$ qubits etc. (without considering the extra diagonal gates) for a total of approximately $2^{n}$ CNOTs and $2^n$ generic one-qubit gates. Finally, to have the total count for the number of elementary rotations, we decompose each one-qubit gate $U$ as a product of three elementary rotations (ignoring the global phase) \cite{nielsen2011quantum}
\begin{equation}
 U = R_x(\alpha) \times R_z(\beta) \times R_x(\gamma) 
 \label{eq::decomp}
 \end{equation}
where $\alpha, \beta, \gamma$ are three real parameters. $R_x$
rotations commute with the CNOT gate if the $R_x$ gate acts on the
target qubit of the CNOT gate. So for each quantum subcircuit
implementing an $SU(2)$-multiplexor and starting from the leftmost
rotation, we can commute the $R_x$ gate,
merge it with the next generic one-qubit
gate, and repeat the process (decomposition shown in Eq. \eqref{eq::decomp}, commutation and merging) until we reach the last one-qubit gate of the multiplexor implementation. Thus, up to a linear number of gates, all the generic one-qubit gates can be decomposed into only two elementary rotations, for a total of approximately $2^{n+1}$ rotation gates.

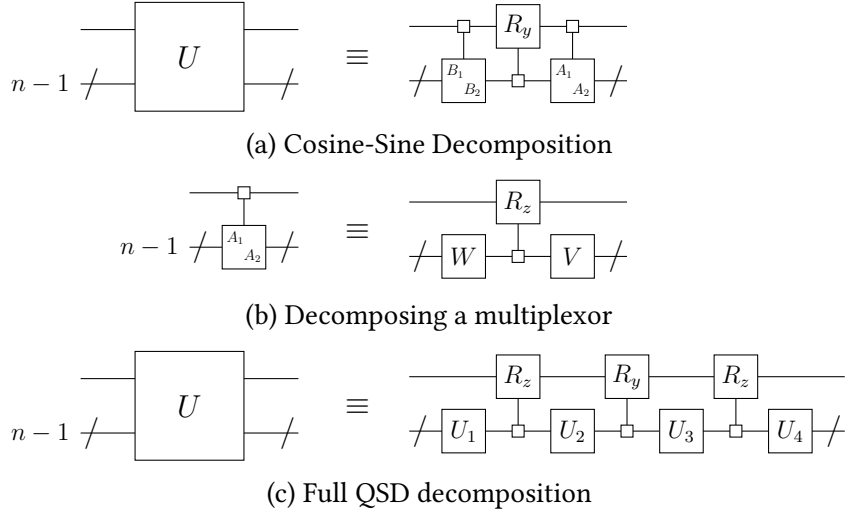
\begin{figure*}[t!]
    \centering
    \begin{subfigure}[t]{0.9\textwidth}
      \centering
\newcommand{\xs}{1}\newcommand{\gs}{.4}\newcommand{\bs}{.11}\newcommand{\nots}{.2}
\newcommand{\bgate}[4]{
  \draw (#1,1) -- (#1,2) ; 
  \draw[fill=white] (#1-\bs,#2-\bs) rectangle (#1+\bs,#2+\bs) ; 
  \draw[fill=white] (#1-\gs,#3-\gs) rectangle (#1+\gs,#3+\gs); 
  \node at (#1,#3) {\scalebox{.8}{#4}} ;}%
\newcommand{\gate}[3]{
  \draw[fill=white] (#1-\gs,#2-\gs) rectangle (#1+\gs,#2+\gs); 
  \node at (#1,#2) {\scalebox{.8}{#3}} ;}%
\begin{minipage}{5cm}
  \hfill
\begin{tikzpicture}[x=4ex,y=-4ex]
\node at (0.2,2) {$/$} ;
\node at (3.8,2) {$/$} ;
\draw (0,1) -- (4,1) ;
\draw (0,2) node [anchor=east] {\scalebox{.8}{$n-1$}} -- (4,2) ;
\draw[fill=white] (1,.5) rectangle (3,2.5) ;
\node at (2,1.5) {$U$} ;
\end{tikzpicture}
\end{minipage}
\quad$\equiv$\quad
\begin{minipage}{7cm}
\begin{tikzpicture}[x=4ex,y=-4ex]
\node at (0.2,2) {$/$} ;
\node at (3.8,2) {$/$} ;
\draw (0,1) -- (4,1) ;
\draw (0,2) -- (4,2) ;
\bgate{1}{1}{2}{\scalebox{.6}{$\begin{array}{c@{}c}B_1&\\[0ex]&B_2\end{array}$}}
\bgate{2}{2}{1}{$R_y$}
\bgate{3}{1}{2}{\scalebox{.6}{$\begin{array}{c@{}c}A_1&\\[0ex]&A_2\end{array}$}}
\end{tikzpicture}
\end{minipage}
        \caption{Cosine-Sine Decomposition}
        \label{qsd1}
    \end{subfigure}\\[1ex]

    \begin{subfigure}[t]{0.9\textwidth} 
        \centering
\newcommand{\xs}{1}\newcommand{\gs}{.4}\newcommand{\bs}{.11}\newcommand{\nots}{.2}
\newcommand{\bgate}[4]{
  \draw (#1,1) -- (#1,2) ; 
  \draw[fill=white] (#1-\bs,#2-\bs) rectangle (#1+\bs,#2+\bs) ; 
  \draw[fill=white] (#1-\gs,#3-\gs) rectangle (#1+\gs,#3+\gs); 
  \node at (#1,#3) {\scalebox{.8}{#4}} ;}%
\newcommand{\gate}[3]{
  \draw[fill=white] (#1-\gs,#2-\gs) rectangle (#1+\gs,#2+\gs); 
  \node at (#1,#2) {\scalebox{.8}{#3}} ;}%
\begin{minipage}{5cm}
  \hfill
\begin{tikzpicture}[x=4ex,y=-4ex]
\node at (0.2,2) {$/$} ;
\node at (1.8,2) {$/$} ;
\draw (0,1) -- (2,1) ;
\draw (0,2) node [anchor=east] {\scalebox{.8}{$n-1$}} -- (2,2) ;
\bgate{1}{1}{2}{\scalebox{.6}{$\begin{array}{c@{}c}A_1&\\[0ex]&A_2\end{array}$}}
\end{tikzpicture}
\end{minipage}
\quad$\equiv$\quad
\begin{minipage}{7cm}
\begin{tikzpicture}[x=4ex,y=-4ex]
\node at (0.2,2) {$/$} ;
\node at (3.8,2) {$/$} ;
\draw (0,1) -- (4,1) ;
\draw (0,2) -- (4,2) ;
\gate{1}{2}{$W$}
\bgate{2}{2}{1}{$R_z$}
\gate{3}{2}{$V$}
\end{tikzpicture}
\end{minipage}
        \caption{Decomposing a multiplexor}
        \label{qsd2}
    \end{subfigure}\\[1ex]

    \begin{subfigure}[t]{0.9\textwidth}
      \centering
\newcommand{\xs}{1}\newcommand{\gs}{.4}\newcommand{\bs}{.11}\newcommand{\nots}{.2}
\newcommand{\bgate}[4]{
  \draw (#1,1) -- (#1,2) ; 
  \draw[fill=white] (#1-\bs,#2-\bs) rectangle (#1+\bs,#2+\bs) ; 
  \draw[fill=white] (#1-\gs,#3-\gs) rectangle (#1+\gs,#3+\gs); 
  \node at (#1,#3) {\scalebox{.8}{#4}} ;}%
\newcommand{\gate}[3]{
  \draw[fill=white] (#1-\gs,#2-\gs) rectangle (#1+\gs,#2+\gs); 
  \node at (#1,#2) {\scalebox{.8}{#3}} ;}%
\begin{minipage}{5cm}
\hfill
  \begin{tikzpicture}[x=4ex,y=-4ex]
\node at (0.2,2) {$/$} ;
\node at (3.8,2) {$/$} ;
\draw (0,1) -- (4,1) ;
\draw (0,2) node [anchor=east] {\scalebox{.8}{$n-1$}} -- (4,2) ;
\draw[fill=white] (1,.5) rectangle (3,2.5) ;
\node at (2,1.5) {$U$} ;
\end{tikzpicture}
\end{minipage}
\quad$\equiv$\quad
\begin{minipage}{7cm}
\begin{tikzpicture}[x=4ex,y=-4ex]
\node at (0.2,2) {$/$} ;
\node at (7.8,2) {$/$} ;
\draw (0,1) -- (8,1) ;
\draw (0,2) -- (8,2) ;
\gate{1}{2}{$U_1$}
\bgate{2}{2}{1}{$R_z$}
\gate{3}{2}{$U_2$}
\bgate{4}{2}{1}{$R_y$}
\gate{5}{2}{$U_3$}
\bgate{6}{2}{1}{$R_z$}
\gate{7}{2}{$U_4$}
\end{tikzpicture}
\end{minipage}
        \caption{Full QSD decomposition}
        \label{qsd3}
    \end{subfigure}
    \caption{Circuit equivalences for the QSD}
    \label{qsd}
\end{figure*}

\medskip
\noindent
\paragraph{Quantum Shannon Decomposition}
Among the various existing synthesis methods
\cite{PhysRevA.94.052317, nielsen2011quantum, PhysRevLett.92.177902},
the one giving the shortest circuits in terms of number of gates is the Quantum
Shannon Decomposition (QSD) \cite{mottonen2005decompositions, 1629135}. It
relies on the following two decomposition formulas: 
\begin{itemize}
\item the first one is the Cosine-Sine decomposition (CSD) of a
  unitary matrix $U$ on $n$ qubits~\cite{GVL96}:
\begin{equation} U = \begin{pmatrix} A_1 & \\ & A_2 \end{pmatrix} \begin{pmatrix} C
    & -S \\ S & C \end{pmatrix} \begin{pmatrix} B_1 & \\ &
    B_2 \end{pmatrix}. \label{CSD} \end{equation}
$A_1, A_2, B_1, B_2$ are unitary matrices on $n-1$ qubits and $C,S$
are real positive diagonal matrices such that
$C^2 + S^2 = I_{2^{n-1}}$. The second term in the CS decomposition
is in fact an $R_y$-multiplexor controlled by the $n-1$ least
significative qubits. The circuit equivalence is given in Figure
\ref{qsd}\subref{qsd1}, where angles are omitted for legibility.
\item The second formula decomposes a multiplexor
\begin{equation} \begin{pmatrix} A_1 & \\ & A_2 \end{pmatrix} = \begin{pmatrix} V &
    \\ & V \end{pmatrix} \begin{pmatrix} D^{\dag} & \\ &
    D \end{pmatrix} \begin{pmatrix} W & \\ & W \end{pmatrix} \label{demultiplexe} \end{equation} with
$D$ a diagonal matrix on $n-1$ qubits, $V$ and $W$ are unitary operating $n-1$ qubits. The second term involving the matrix $D$ is in fact a 
$R_z$ multiplexor controlled by the $n-1$ least significative
qubits. The circuit equivalence is represented in Figure
\ref{qsd}\subref{qsd2}, again with omitted angles.
\end{itemize}~\\
Finally, synthesizing $U$ on $n$ qubits is equivalent to synthesizing 3 rotation
multiplexors on $n$ qubits and 4 matrices on $n-1$ qubits on which we
can apply the QSD again as showed in Figure \ref{qsd}\subref{qsd3}. We repeat the process until we get only
multiplexors and gates acting on a small number of qubits (typically
2) for which an exact decomposition is known
\cite{Bullock:2003:ATC:775832.775916, PhysRevA.63.062309, PhysRevA.69.032315, vatan2004realization, PhysRevA.69.010301}.

If the Quantum Shannon Decomposition gives the best asymptotic number
of CNOTs in the circuit: $\frac{23}{48} \times 4^n$, this method has
nonetheless drawbacks. It does not take into account other metrics
useful to minimize in the compilation process such as the classical
time required to compute the circuit. The algorithm for computing Formula~\eqref{CSD} 
consists in reducing the $K \times K$ matrix $U$ into a $2 \times 2$ bidiagonal block form, then the 4 bidiagonal blocks are simultaneously diagonalized using bidiagonal SVD algorithms. The first part is the most expensive one in terms of floating point operations: by applying Householder reflectors to the left and right of $U$, we progressively bidiagonalize $U$ --- this requires $K^3/3$ flops for each block --- and we store the accumulation of each Householder reflector to compute $A_1, A_2, B_1, B_2$ --- this requires $K^3/6$ flops for each block. Overall, computing the CSD on a $K \times K$ matrix requires $2 \times K^3$ flops
\cite{Sutton2009}. Concerning Formula~\eqref{demultiplexe}, one has to perform two matrix/matrix products and an eigenvalue decomposition. With square matrices of size K, each matrix/matrix product on matrices requires $2 \times K^3$ flops and the eigenvalue decomposition needs around $26 \times K^3$ \cite[Table 3.13]{BenchLAPACK}. Overall for the first step of the Quantum Shannon Decomposition of a matrix of size $N$ we have to compute one CSD of a matrix of size $N$ and decompose two multiplexors i.e four matrix/matrix products of size $N/2$ and two eigenvalue decompositions of size $N/2$ too. This represents a total of $2 \times N^3 + 4 \times 2 \times (N/2)^3 + 2 \times 26 \times (N/2)^3 = \frac{19}{2} \times N^3$ flops. Then to pursue the algorithm we have to perform the same operations on 4 matrices of size $N/2$, then on 16 matrices of size $N/4$ etc. Overall, with $N = 2^n$ we can approximate the total number of flops to $19 \times 8^n$ which is very expensive. In the next section we propose
an alternative method based on Householder transformations. Strongly
connected to classical results about QR decomposition, this method
aims at achieving better performance in the synthesis of quantum
circuits by finding a compromise between circuit size and calculation
time.

\section{Householder algorithm for unitary matrices}\label{sec:householder}

In this section, we first recall the main principles of the QR factorization of a general complex square matrix via Householder transformations. Then we consider the special case of unitary matrices that correspond to quantum operators.

The QR decomposition of a matrix $A \in \mathbb{C}^{n \times n}$ expresses $A$ as the product of a unitary matrix $Q \in \mathcal{U}(n)$ and an upper triangular matrix $R$. A standard algorithm to compute such a factorization consists in applying a series of Householder transformations~\cite[p. 209]{GVL96} zeroing out successively the subdiagonal entries of each column. 

\begin{figure}[t]
  \centering
  $A^{(k)}\quad=\quad
  \begin{array}{c}\begin{tikzpicture}[x=3ex,y=-3ex]
    \draw[thick] (0,0) -- (0,6) -- (6,6) -- (6,0) -- (0,0) ;
    \draw (0,0) -- (2,2) -- (6,2) ;
    \draw (2,2) -- (2,6) ;
    \node at (1,3.5) {$0_{k-1}$} ;
    \node at (3.5,1) {$R^{(k)}$} ;
    \node at (4.25,4) {$B^{(k)}$} ;
    \node[color=blue] at (2.5,4) {$b$} ;
    \draw[color=blue, thick] (2.5,4) ellipse (.4 and 1.8) ;
  \end{tikzpicture}
  \end{array}$
\caption{Matrix pattern at step $k$-th of Householder transformation}
\label{House}
\end{figure}
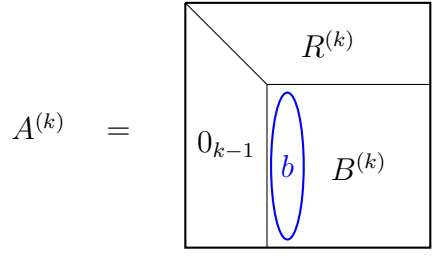

At step $k$ ($1 \leq k \leq n-1$) of the QR algorithm, we zero out all but the first entry of the vector $b$ in the matrix depicted in Figure~\ref{House} using the Householder transformation,   
$H'_k = I_n - \tau_k u_k u_k^{\dag}$,
where $u_k \in \mathbb{C}^{n-k+1}$ and $\tau_k=2/u_k^{\dag}u_k$. Note that in the complex case, the Householder matrix $H'_k$ can be sometimes referred to as ``elementary unitary matrix'' 
(e.g. in~\cite{Lehoucq:1996:CEU:235815.235817}).

Then the $k$-th iteration ends with the computation of the matrix 
\[ A^{(k)} = H_k A^{(k-1)} ,\]
with $H_k = \begin{pmatrix} I_{k-1} && 0 \\ 0 && H'_k \end{pmatrix}$, $A^{(0)}=A$ and $H_1=H'_1$.
This operation updates $B^{(k)}=A^{(k)}(k:n,k:n)$ via the relation
\begin{equation}\label{updateB}
\left(I_{n-k+1} - \tau_k u_k u_k^{\dag} \right) B^{(k)} = B^{(k)} - \tau_k u_k \left(\left(B^{(k)}\right)^{\dag} u_k \right)^{\dag},
\end{equation} which zeroes out the subdiagonal entries in column $k$ (but does not affect the zeros already introduced in previous columns). 
The problem is now to find the vector $u_k \in \mathbb{C}^{n-k+1}$ such that 
\[ \left(I_{n-k+1} - \tau_k u_k u_k^{\dag} \right) b  = \left(\beta_k, 0, \\ \hdots, 0 \right)^T = \beta_k e_1. \]
with $\beta_k \in \mathbb{C}$. 
From~\cite[p. 233]{GVL96}, we have $u_k=b \pm e^{i\theta} \|b \| e_1$ with $\theta = \arg(b_1) $but various choices for $u_k$ have been proposed in numerical libraries (see~\cite{Lehoucq:1996:CEU:235815.235817} for a review of these choices).
 
At the end of the algorithm we have computed a set of $n-1$ Householder transformations $H_1, H_2, ..., H_{n-1}$ such that
\[ \left(\prod_{i=1}^{n-1} H_{n-i}\right) A = R \]
where $R$ is upper triangular. Since the Householder matrices are Hermitian, we obtain  
\[ A = \left(\prod_{i=1}^{n-1} H_i\right) R = QR. \]
In the QR algorithm, the Householder matrices $H'_k$ never need to be explicitly formed and the expensive part of the computation is the update of the matrix  $B^{(k)}$, given in 
Equation~(\ref{updateB}), which requires at each iteration a matrix-vector multiplication followed by a rank-1 update of $B^{(k)}$. The total cost of the factorization is about $\frac{4}{3} n^3$ complex flops ($\frac{16}{3} n^3$ real flops).

A block version of the algorithm uses the fact that a product of $p$ Householder matrices $H_1\times \hdots \times H_p$ can be written as $I - VTV^{\dag}$ where $V$ is an $n \times p$ rectangular matrix with the Householder vector $u_k \in \mathbb{C}^n$ at the $k$-th column and $T$ is an upper triangular matrix \cite{doi:10.1137/S0895479894276369}. The algorithm consists in partitioning $A$ into blocks of size $n \times n_A$ for some $n_A$, factorizing the first block and updating the remaining blocks via the operation (we use a matlab-like notation)
\[ A(:,n_A+1:n) = A(:,n_A+1:n) - VTV^{\dag} A(:,n_A+1:n), \]
and repeating the process with the next block until the whole matrix
is triangularized. The update is richer in BLAS 3 operations~\cite{BLAS}, potentially leading to better performance, yet without decreasing the flop count~\cite{doi:10.1137/0910005}.

\medskip

Let now exploit the specificity of quantum operators where the corresponding matrix $A$ is unitary and see how the QR decomposition simplifies. In this case, the triangular factor is also unitary and thus diagonal and the QR algorithm of $A$ consists in a progressive diagonalization of $A$. 
For sake of simplification, we detail in the remainder only the first iteration.
Let $b=(b_1,\hdots,b_{n})^T$ be the first column of the unitary matrix $A$ and $r = (r_1,\hdots,r_{n})$ its first row. 
We choose the value of the Householder vector $u=b \pm e^{i\theta} \|b \| e_1$ as defined 
in~\cite[p. 233]{GVL96} but we will choose the sign ``+''.
This choice has the advantage of maximizing $\|u \|$ (for sake of stability~\cite[p. 233]{GVL96}) and of simplifying the final decomposition of the quantum operator into elementary circuits, as we will see in Section~\ref{sec:circuit}. 
Since $A$ (and $A^{(k)}$ at the $k$th iteration) is unitary, we have $\|b\|=1$ and we get 
\[ u = b + e^{i\theta} e_1 \]
and 
\[ \tau = \frac{2}{\|u\|^2} = \frac{2}{\|b\|^2 + \|e_1\|^2 + 2|b_1|} = \frac{1}{1 + |b_1|}. \]
Then applying the Householder transformation $H$ to $b$ gives 
\begin{equation} \label{householder} H b = -e^{i\theta} \|b \| e_1 = -e^{i\theta} e_1 . \end{equation}
The gain in complexity occurs in the update phase. Using the orthonormality of the vectors of $A$, the update expressed in Equation~(\ref{updateB}) simplifies to
\[ H A = A - \tau (b + e^{i\theta} e_1)\left(A^{\dag}\left(b + e^{i\theta} e_1\right)\right)^{\dag} = A - \tau (b + e^{i\theta}e_1)(e_1^T + e^{-i\theta} r).  \]
Then we have
\[ H A = A - \tau (b e_1^T + e^{i\theta}e_1 e_1^T + e_1 r + e^{-i\theta} b r).\]
The first column of $A$ does not need to be updated in this computation because of Equation~\eqref{householder} then we can ignore the term $b e_1^T + e^{i\theta}e_1 e_1^T$.
Similarly the first row of $A$ does not need to be updated because the unitarity of the rows of $A$ ensures that $r = -e^{i\theta} e_1^T$ after application of $H$. 
Moreover $\tau e^{-i\theta} = 1/(e^{i\theta} + |b_1|e^{i\theta})=1/(b_1 + e^{i\theta})=1/u_1$, where $u_1$ denotes the first component of $u$.
So we are left with the rank-1 update
\[\left(H A\right)_{2:n,2:n} = A_{2:n,2:n} - \frac{b(2:n) \; \cdot \; r(2:n)}{u_1} . \]
The matrix-vector product expressed in Equation~(\ref{updateB}) for the classical QR factorization is avoided. The matrix obtained after the first iteration is then 
\[ A^{(1)} = \begin{pmatrix} -e^{i\theta} && 0 \\ 0 && \left(H A\right)_{2:n,2:n}\end{pmatrix} \]
and we can continue the algorithm on the unitary matrix $A^{(1)}(2:n,2:n)$ and so on, until $A$ becomes diagonal. The update at the k-th iteration requires only $(n-k)^2$ multiplications and $(n-k)^2$ additions. Finally this new algorithm requires $\sum_{k=1}^{n-1} 2 \times (n-k)^2 \sim \frac{2}{3} n^3$ complex flops, which is twice as less than the standard case.

It is possible to choose the vector $u$ such that $u_1=1$, then the value of $\tau$ will be adjusted so that the resulting Householder transformation $H$ remains the same. More precisely, keeping the notations above we set 
\begin{equation} u \leftarrow \frac{1}{e^{i\theta}(1+|b_1|)} u \label{normalization} \end{equation}
and we obtain 
$\tau= (1+|b_1|)$ and then the update phase becomes
\begin{equation} \left(H A\right)_{2:n,2:n} = A_{2:n,2:n} - u(2:n) \; \cdot \; r(2:n) . \label{rank-one} \end{equation}
The algorithm can easily be done in place. One can store the Householder vectors in the strictly lower triangular part of $A$, the diagonal elements of $R$ are stored in the diagonal and the $\tau_i$'s are stored in a specific array. 

The main cost of the algorithm resides in the rank-one update phase
in Equation~\eqref{rank-one}. 
In order to use more optimized BLAS 2 and BLAS 3 operations we can derive from Equation \eqref{rank-one} new update relations. Suppose we have already performed the factorization and the update for  the first $nb$ rows and columns for some $nb$. Therefore the first $nb$ columns of $A$ contain the Householder vectors, and the block $A(1:nb, nb+1:n)$ has been updated following \eqref{rank-one}. Let $i,j \in \llbracket nb+1, n \rrbracket$, one can verify that the update of the element $A(i,j)$ is given by  
\begin{equation} A(i,j) \leftarrow A(i,j) - \sum_{k=1}^{nb} A(i,k) \times A(k,j)  \label{single-update} \end{equation}
by simply applying successively the update \eqref{rank-one}.

In terms of matrix and vector operations we have 
\begin{equation} A(i,nb+1:n) \leftarrow A(i,nb+1:n) - A(i, 1:nb) \times A(1:nb, nb+1:n) \label{row-update} \end{equation}
for the update of one row, 
\begin{equation} A(nb+1:n, j) \leftarrow A(nb+1:n,j) - A(nb+1:n, 1:nb) \times A(1:nb, j) \label{column-update} \end{equation}
for the update of one column and 
\begin{multline} A(nb+1:n, nb+1:n) \leftarrow\\ A(nb+1:n, nb+1:n) - A(nb+1:n, 1:nb)\times A(1:nb, nb+1:n) \label{matrix-update} \end{multline}
for the update of the full matrix. This last update is a BLAS 3
operation and can potentially yield higher performance on hybrid
CPU-GPU architectures~\cite{TDB.10}.

\medskip
Using these new update relations we can improve the algorithm by three means: 
\begin{itemize}
\item first we can improve the unblocked algorithm. Instead of updating the whole matrix at each iteration with a rank one update we only update one row and one column : at the k-th iteration we have computed the k-th Householder vector and we update the row $A(k+1, k+1:n)$ and the column $A(k+1:n, k+1)$ via the relations \eqref{row-update} and \eqref{column-update}. Such updates consist in more and bigger matrix-vector operations and experimentally it appears to scale better. 
\item Secondly this naturally leads to a blocked version of the
  algorithm. Let $nb$ be the size of our block. Once we have done the computations on the first $nb$ rows and $nb$ columns of $A$ with an unblocked version, we can update the rest of the matrix with a matrix/matrix product via equation \eqref{matrix-update} and continue the algorithm on the matrix $A(nb+1:n, nb+1:n)$ until we reach the last block where the unblocked algorithm is applied. 
\item A third improvement can be made in order to avoid using the unblocked algorithm to compute the full panel of $nb$ rows and columns of $A$. Indeed the update of the blocks $A(nb+1:n, 1:nb)$ and $A(1:nb, nb+1:n)$ can be performed with BLAS 3 operations. One can prove that there exist triangular matrices $T_1^i, T_2^i$ of size $i \times i, i=1..nb$ such that 
\begin{equation} A(i+1:n, 1:i) \leftarrow A(i+1:n, 1:i) \times T_1^i \label{update-blockcolumns} \end{equation}
\begin{equation} A(1:i, i+1:n) \leftarrow T_2^i \times A(1:i, i+1:n). \label{update-blockrows} \end{equation}
The matrices $T_1^i, T_2^i$ are computed using the following recursive formula: 
\begin{equation} T_1^{1} = 1, T_1^{i+1} = \begin{pmatrix} T_1^{i} & -p_{i+1}T_1^{i}/\mathcal{N}_i \\ & 1/\mathcal{N}_i \end{pmatrix}, \end{equation}
\begin{equation} T_2^{1} = 1, T_2^{i+1} = \begin{pmatrix} T_2^{i} &  \\ -q_{i+1}T_2^{i}  &1\end{pmatrix}\label{lastformula} \end{equation}
with $p_i = A(1:i, i), q_i = A(i,1:i)$. $\mathcal{N}_i$ is the normalization factor expressed in Equation~\eqref{normalization}.

\begin{proofof}{Proof of Formulas~\eqref{update-blockcolumns} to~\eqref{lastformula}} 
By induction on $i$. We do it for $T_2$ only. The case $i=1$ is trivial because we do not have to update the first row. Now suppose the result is true for some $i$, $1 \leq i < nb$. The first $i$ rows are already updated by the application of $T_2^i$, we only need to update the next row $i+1$. Let $A(i+1,j), j \in \llbracket nb+1,n \rrbracket$ be an element of this row. If the column $A(1:i,j)$ was already updated the update of $A(i+1,j)$ would be given by the equation \eqref{single-update} i.e 
\[ A(i+1,j) \leftarrow A(i+1,j) - \sum_{k=1}^i A(i+1,k) \times A(k,j). \] 

Written differently $A(i+1,j) \leftarrow A(i+1,j) - A(i+1,1:i+1) \cdot A(1:i,j)$. By hypothesis $A(1:i,j)$ is updated by the relation $A(1:i,j) \leftarrow T_2^{i} A(1:i,j)$. This gives 
\[ A(i+1,j) \leftarrow [-A(i+1, 1:i+1) \times T_2^{i} \; ; \; 1]  \cdot A(1:i+1,j) . \] 

Doing it for all $j$ and concatenating it with the update of the first $i$ rows by the action of $T_2^{i}$ gives the result. The same thing can be done with $T_1$ but one has to be careful about the normalization of the Householder vectors. 
\qedhere
\end{proofof}

\end{itemize}

Therefore $T_1^{nb}$ and $T_2^{nb}$ only depend on the block $A(1:nb, 1:nb)$ and can be used to update $A(nb+1:n, 1:nb)$ and $A(1:nb, nb+1)$ in two BLAS 3 updates. This means that during an iteration we only need to perform an unblocked Householder factorization on a square matrix of size $nb$ and then perform 3 BLAS 3 updates. The pseudo code of the algorithm is given in Algorithm~\ref{PseudoCode} (we call the corresponding routine ZUNQRF and its unblocked version ZUNQR2). ZLARFT2 refers to the adaptation of the standard ZLARFT routine that computes the triangular matrices. 

\begin{algorithm}
\caption{Householder factorization of a unitary matrix $A$ - ZUNQRF}
\label{PseudoCode}
\begin{algorithmic} 
\REQUIRE $N \geq 0, \; \; $A$ \in \mathcal{U}_N$
\ENSURE $A = QR$
\STATE // $NX$ determines when to switch from blocked to unblocked code
\STATE // $NB$ is the block size
\FOR{$I = 1, NX, NB$}
\STATE $IB \leftarrow MIN(N-I+1, NB)$
\STATE call ZUNQR2( IB, IB, A( I, I ), TAU( I )  )
\STATE $T_1, T_2 \leftarrow$ ZLARFT2( N, IB, A( I, I ), TAU( I )  )
\STATE update $A(I:N, I:I+IB)$ via a call to ZTRMM
\IF{$I+IB \leq N$}
\STATE update $A(I:I+IB, I:N)$ via a call to ZTRMM
\STATE update $A(I+IB:N, I+IB:N)$ via a call to ZGEMM
\ENDIF
\ENDFOR
\IF{$I \leq N$}
\STATE call ZUNQR2( N-I+1, N-I+1, A( I, I ), TAU( I )  )
\ENDIF
\end{algorithmic}
\end{algorithm}

Thanks to the above QR decomposition resulting in a product of Householder matrices and a diagonal matrix, we store the information of a unitary matrix into the subdiagonal part of the complex matrix (the Householder vectors), and two real vectors, including the $\theta$'s (angles of the diagonal entries) and the $\tau$'s. 
In the next section we use this factorization of unitary operators to obtain quantum circuits .

\section{From the Householder decomposition to a quantum circuit}\label{sec:circuit}
In this section we develop several methods to convert the Householder representation of a unitary matrix into a quantum circuit. We present a general method in the Section~\ref{sec:4.1} and we optimize it in Section~\ref{sec:optim}. 

\subsection{General method}\label{sec:4.1}
Let $U \in U(2^n)$ be the unitary matrix we want to synthesize. The QR
factorization of $U$ gives normalized vectors $u_1, u_2, \hdots, u_{2^n-1}$ and a diagonal matrix $D$ such that
\[ U = \prod_{i=1}^{2^n-1} H_{i} \times D. \]
where $H_i$ are Householder matrices defined by
$ H_i = I_{2^n} - 2u_iu_i^\dagger $
as in Section~\ref{sec:householder} (since the $u_i$ are normalized).
The synthesis of a diagonal operator is a well-known
problem~\cite{Bullock:2004:AOC:2011572.2011575, welch2015synthesis}. Therefore, the main issue is the synthesis of the Householder matrices. We recall that 
\[ H_i u_i = -u_i \]
and
\[ \forall v, v \perp u_i \Rightarrow H_i v = v. \]
Consequently, for any unitary matrix $P_i$ whose first column is $u_i$
we can write
\begin{equation} \label{householder_decomposition} H_i = P_iD_GP_i^{\dag} \end{equation}
with
\[ D_G = \begin{pmatrix} -1 && && && \\ && 1 && && \\ && && \ddots && \\
    && && && 1 \end{pmatrix}. \]
Indeed $P_i$ can be regarded as an orthonormal basis of vector columns
containing $u_i$. In other words,
Equation~\eqref{householder_decomposition} is a diagonalization of $H_i$.

\begin{figure}[t]
  \centering
  \newcommand{\myxxh}{2}\newcommand{\myxxw}{.5}\newcommand{\myxxx}[3]{
    \draw[fill=white] (#1-\myxxw,#2-\myxxh) rectangle (#1+\myxxw,#2+\myxxh) ;
    \node at (#1,#2) {\scalebox{.7}{$#3$}} ;
  }
  \newcommand{\myyyh}{.8}\newcommand{\myyyw}{1.5}\newcommand{\myyy}[3]{
    \draw[fill=white] (#1-\myyyw,#2-\myyyh) rectangle (#1+\myyyw,#2+\myyyh) ;
    \node at (#1,#2) {\scalebox{.7}{$#3$}} ;
  }
  \scalebox{.95}{
    \begin{tabular}{c}\begin{tikzpicture}[x=3ex,y=-3ex]
        \draw (0,0) -- (4,0) ;
        \node at (0,0) [anchor=south east] {\scalebox{0.8}{$|0\rangle^n$}} ;
        \node at (0.2,0) {$/$} ;
        \draw[fill=white] (1,-1) rectangle (3,1) ;
        \node at (2,0) {$U$} ;
      \end{tikzpicture}\end{tabular}
    $\equiv$
    \begin{tabular}{c}\begin{tikzpicture}[x=3ex,y=-3ex]
        \draw (0,0) -- (11,0) ;
        \node at (0.2,0) {$/$} ;
        \myxxx{1.5}{0}{D}
        \myyy{4}{0}{\textit{SP}(u_{2^n-1})^{\dagger}}
        \myxxx{6.5}{0}{D_G}
        \myyy{9}{0}{\textit{SP}(u_{2^n-1})}
        \node at (12,0) {$\cdots$} ;
        \draw (13,0) -- (22,0) ;
        \myyy{15}{0}{\textit{SP}(u_{1})^{\dagger}}
        \myxxx{17.5}{0}{D_G}
        \myyy{20}{0}{\textit{SP}(u_1)}
      \end{tikzpicture}\end{tabular}
    }
\caption{A first circuit for the Householder method}
\label{Householder_SP}
\end{figure}
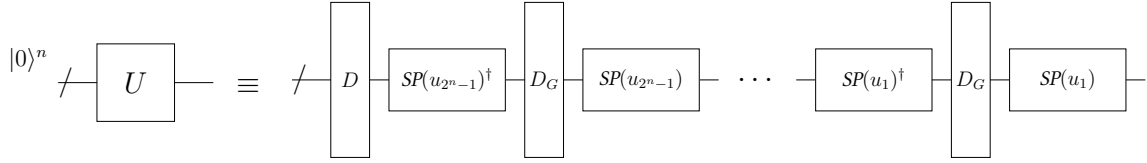

From this analysis, we get the following decomposition of the unitary matrix $U$
\[ U = \prod_{i=1}^{2^n-1} P_i D_G P_i^{\dag} \times D\]
and we can derive a quantum circuit as depicted in
Figure~\ref{Householder_SP}.
\begin{itemize}
\item As mentionned already, the synthesis of $D$ is a problem with
  known solutions.
\item Each block $P_i D_G P_i^{\dag}$ is equivalent to de-preparing
  the state $u_i$, applying $D_G$ and reprepar\-ing the state $u_i$. 
  \begin{itemize}
  \item The matrix $D_G$ is the ``zero phase shift'' operator and is
    used for instance in the Grover diffusion operator in Grover's
    algorithm \cite{Grover:1996:FQM:237814.237866}. 
  \item In our circuits we use the notation $\textit{SP}(v)$ to refer to a black box
    that prepares the state $v$. Although many different operators can
    prepare the state $v$, we point out that in one circuit the
    operators preparing and de-preparing the same state are exactly the
    same, otherwise the decomposition would not be valid.
    Many previous research studies have sought to optimize the preparation of
    states and we use their results for our synthesis
    \cite{mottonen2005decompositions, PhysRevA.83.032302, 1629135}.
  \end{itemize}
\end{itemize}

\subsection{Resources estimation}\label{sec:optim}
We now turn to the question of the size of the circuit sketched in
Figure~\ref{Householder_SP} (measured in number of CNOTs), and to the
computational cost to generate the circuit (measured in
flops). In this section we only give asymptotic results which are
summarized in Tables~\ref{gate_count} and~\ref{flop_count}. 

Asymptotically, it turns out that the synthesis of $D$ and and $D_G$
are negligible. Using existing methods one can synthetize $D$ in
$O(2^n)$ gates, while the series of $2^{n}-1$ subcircuits $D_G$
requires at most
$O((2^n-1)n^2)$ gates~\cite{PhysRevA.93.032318}
(when $n$ is the number of qubits). As we will
see in the following, the synthesis of the $P_i$'s requires $O(4^n)$ gates: 
this is the dominant factor.

The complexity of the size of the circuit is therefore essentially due to the
preparation and de-preparation of quantum states. Because of the
structure of the problem, we can do better than systematically applying state
preparation on $n$ qubits for each of the $u_i$ vectors. We
describe two successive optimizations.
The first one relies on the possibility to perform state preparations
on less than $n$ qubits;
the second one proposes to fuse adjacent sequences of
de-preparations and preparation of states. 

\subsubsection{Optimization based on state preparation}
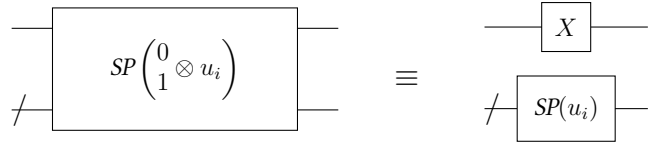
\begin{figure}[t]
  \centering
    \begin{tabular}{c}\begin{tikzpicture}[x=3ex,y=-3ex]
        \draw (0,0) -- (8,0) ;
        \draw (0,2) -- (8,2) ;
        \node at (0.2,2) {$/$} ;
        \draw[fill=white] (1,-.5) rectangle (7,2.5) ;
        \node at (4,1) {\scalebox{.8}{$\textit{SP}\left(\begin{matrix}0\\1\end{matrix}\otimes{u_i}\right)$}} ;
      \end{tikzpicture}\end{tabular}
    \quad
    $\equiv$
    \quad
    \begin{tabular}{c}\begin{tikzpicture}[x=3ex,y=-3ex]
        \draw (0,0) -- (4,0) ;
        \draw (0,2) -- (4,2) ;
        \node at (0.2,2) {$/$} ;
        \draw[fill=white] (1.4,-.6) rectangle (2.6,.6) ;
        \node at (2,0) {\scalebox{.8}{$X$}} ;
        \draw[fill=white] (.8,2-.8) rectangle (3.2,2+.8) ;
        \node at (2,2) {\scalebox{.8}{$\textit{SP}({u_i})$}} ;
      \end{tikzpicture}
    \end{tabular}
    \caption{Desentangling one qubit in state preparation}
\label{SP_X}
\end{figure}

When preparing $u_i$, if only the last $2^k$ elements of
$u_i \in \mathbb{C}^{2^n}$ are non zero, the state is encodable on $k$
qubits only. This means that a k-qubit operator can prepare the state
$u'_i \in \mathbb{C}^{2^k}$ such that
$u_i=(\begin{smallmatrix}0\\u'_i\end{smallmatrix})$. Let $Q$ be such
an operator, then the operator
\[ P = X^{\otimes (n-k)} \otimes Q = \begin{pmatrix} (0) && && 1 \\   && \iddots && \\ 1 && && (0) \end{pmatrix} \otimes \begin{pmatrix} && \\ u'_i  && (*) \\ && \end{pmatrix} = \begin{pmatrix} 0 & * \\ u'_i & * \end{pmatrix} \]
prepares $u_i$. A quantum circuit to illustrate this is given in Figure
\ref{SP_X}.

Thus, up to the operators $X$, $D$ and $D_G$, we observe that the
Householder method breaks down as follows: the synthesis of the first
$2^{n-1}$ columns is done via operators acting on $n$ qubits, then the
next $2^{n-2}$ columns are synthesized with operators acting on $n-1$
qubits, {\em etc}.

In \cite{PhysRevA.93.032318} the concept of isometries is
formalized. Formally, with $n > m$, an $m$ to $n$ qubits isometry can
be represented as a $2^n \times 2^m$ matrix $V$ such that
\[ V^{\dag}V = I_{2^m \times 2^m}. \] The data of the $2^m$ first
columns of an operator on $n$ qubits can be regarded as such an
isometry $V$.  For instance an isometry from 0 to $n$ qubits is a
quantum state on $n$ qubits. Synthesizing an isometry from $n-1$ to
$n$ qubits is
equivalent to synthesizing the first $2^{n-1}$ columns of an n-qubit
operator. With this formalism the synthesis of an $n$-qubit operator via
the Householder method naturally leads to synthesizing $n$ isometries,
respectively isometries from $k-1$ to $k$ qubits for $k$ from $1$ to $n$. Therefore
we introduce the following notations:
\begin{itemize}
\item $h_k$ refers to the number of CNOTs necessary to the synthesis
  of an isometry from $k-1$ to $k$ qubits with the Householder method,
\item $c_n$ refers to the number of CNOTs necessary to the synthesis
  of an $n$-qubit operator with the Householder method,
\end{itemize}
and, refering to the discussion above, we have 
\begin{equation}
  c_n \sim \sum_{k=1}^n h_k.
  \label{eq:cbhk}
\end{equation}
With this decomposition we voluntarily omit the side-effects that may
occur between two subcircuits acting on a different number of qubits
--- typically the transition between the subcircuit preparing states
on $j$ qubits only and the subcircuit preparing states on $j+1$
qubits. These side-effects are asymptotically negligible and not taking them into account highly 
simplifies the calculations.

We are concerned with estimating $c_n$: to this end we focus on the estimation
of $h_k$.
A lower bound to the asymptotic behavior of $h_k$ is given in
\cite{PhysRevA.93.032318}:
\[ h_k^{min} = \frac{3}{16} 4^k + o(4^k).\] With
Equation~\eqref{eq:cbhk} we derive the lower bound $\frac{1}{4} 4^n$
for $c_n$.

With our current circuit, we have 
\begin{equation} 
h_k \sim 2\times(2^{k-1})\times p_k 
\label{eq:hk}
\end{equation} where $p_k$ is the number of
CNOTs required to prepare a state on $k$ qubits. The value $p_k$
varies depending on the
structure of subcircuits we consider
\cite{mottonen2005decompositions}:
\begin{itemize}
	\item with rotation multiplexors, $p_k = 2^{k+1}$, $h_k \sim 2\times4^k$, hence
	\[ c_n \sim \frac{8}{3} 4^n;\]
	\item with multiplexors in $SU(2)$, $p_k = 2^k$, $h_k \sim 4^k$ and 
	\[ c_n \sim \frac{4}{3} 4^n.\]
\end{itemize} 
The same calculation can be done for the number of rotations in the
circuit.
Actually, Equations \eqref{eq:cbhk} and \eqref{eq:hk}
  highlight the decomposition of the quantum circuit into smaller
  subcircuits, thus remain true by replacing the number of CNOTs
  with the number of elementary rotations. A quantum state preparation
  on $n$ qubits requires $2^{n+1}$ rotations, whether using rotations
  or $SU(2)$ multiplexors \cite{mottonen2005decompositions}. Overall
  the number of rotations $r_n$ required for the synthesis of a
  n-qubit operator with the Householder method is
\[ r_n = \frac{8}{3} 4^n. \]

\begin{table}
\centering
\begin{tabular}{lcc} \toprule
    {Method} & {CNOT count} & {Rotation count}  \\ \midrule
    QR & $8.7 \times 4^n$ & Unavailable  \\
    Quantum Shannon & $23/48 \times 4^n$ & $9/8 \times 4^n$   \\ \midrule
    Householder (with rotation multiplexors) & $2 \times 4^n$ & $2 \times 4^n$  \\
    Householder (with multiplexors in $SU(2)$) & $4^n$ &  $2 \times 4^n$  \\ \midrule
    Lower bound & $1/4 \times 4^n $ & $4^n$ \\  \bottomrule
\end{tabular}
\caption{Asymptotic gate counts for decomposition methods}
\label{gate_count}
\end{table}

\subsubsection{Optimizing adjacent state preparations and de-preparations}\label{sec:adjacent}
We now focus on the concatenations of the adjacent subcircuits $\textit{SP}(u_{i+1})$
and $\textit{SP}(u_i)^\dagger$ in the circuit of
Figure~\ref{Householder_SP}. These sequences of operations can indeed
be optimized.

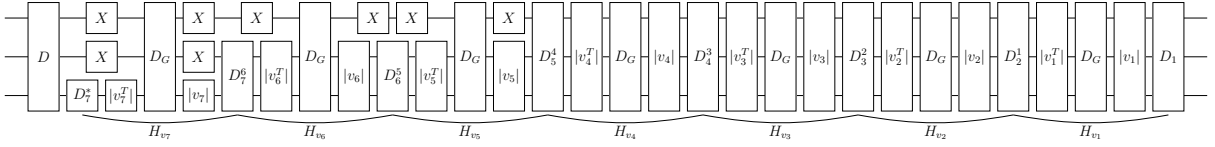
\begin{figure}[t]
  \newcommand{\g}[4]{
    \draw[fill=white] (#1-.4,#2-.4) rectangle (#1+.4,#3+.4) ;
    \pgfmathparse{(#3+#2)/2} \let \z \pgfmathresult
    \node at (#1, \z) {\scalebox{.5}{#4}};
  }
  \centering
  \scalebox{.95}{\begin{tikzpicture}[x=3ex,y=-3ex]
        \draw (0,0) -- (31,0) ;
        \draw (0,1) -- (31,1) ;
        \draw (0,2) -- (31,2) ;
        \g{1}{0}{2}{$D$}
        \g{2.5}{0}{0}{$X$}
        \g{2.5}{1}{1}{$X$}
        \g{2}{2}{2}{$D_7^*$}
        \g{3}{2}{2}{$|v_7^T|$}
        \g{4}{0}{2}{$D_G$}
        \g{5}{0}{0}{$X$}
        \g{5}{1}{1}{$X$}
        \g{5}{2}{2}{$|v_7|$}
        \g{6.5}{0}{0}{$X$}
        \g{6}{1}{2}{$D^6_7$}
        \g{7}{1}{2}{$|v_6^T|$}
        \g{8}{0}{2}{$D_G$}
        \g{9.5}{0}{0}{$X$}
        \g{10.5}{0}{0}{$X$}
        \g{9}{1}{2}{$|v_6|$}
        \g{10}{1}{2}{$D^5_6$}
        \g{11}{1}{2}{$|v_5^T|$}
        \g{12}{0}{2}{$D_G$}
        \g{13}{0}{0}{$X$}
        \g{13}{1}{2}{$|v_5|$}
        \def\gg#1#2#3{\g{#1}{0}{2}{$D^#2_#3$}
          \g{#1+1}{0}{2}{$|v_#2^T|$}
          \g{#1+2}{0}{2}{$D_G$}
          \g{#1+3}{0}{2}{$|v_#2|$}}
        \gg{14}{4}{5}
        \gg{18}{3}{4}
        \gg{22}{2}{3}
        \gg{26}{1}{2}
        \g{30}{0}{2}{$D_1$}
        \def\capt#1#2{\draw (#1,2.5) to[out=-10,in=-170] (#1+4,2.5) ;
          \node at (#1+2,2.5)[anchor=north]{\scalebox{.5}{$H_{v_#2}$}};}
        \capt{2}{7}\capt{6}{6}\capt{10}{5}\capt{14}{4}\capt{18}{3}\capt{22}{2}\capt{26}{1}
      \end{tikzpicture}}
\caption{Quantum circuit designed by the Householder method for 3 qubits}
\label{householder-final-circuit}
\end{figure}

To this end we need to look in more details to the state
preparation circuits. A circuit $P$ that prepares the state $\psi$ on
$n$ qubits can be decomposed as
\[ P = DY \]
where 
\[ D = \begin{pmatrix} e^{i\theta_1} && && \\ && \ddots && \\ && && e^{i\theta_{2^n}} \end{pmatrix} \]
such that $\theta_j = \textit{arg}(\psi(j))$ and $Y$ prepares the real state
\[ \Psi = \begin{pmatrix} |\psi_1| \\ \vdots \\ |\psi_{2^n}| \end{pmatrix}. \]
$Y$ can be synthesized with only $R_y$ rotations by following the
standard methodology for state preparation~\cite{1629135} without caring about the phases equal to 0. Using this decomposition the products preparation/de-preparation that we encounter in the global decomposition are of the form
\[ P_j^{\dag} P_i = Y_j^T D_j^* D_i Y_i  \]
and the diagonal matrices can merge, thus diminishing the size by a cost of a diagonal matrix. In total $2^{k-1}-1$ diagonals on $k$ qubits vanish. Asymptotically this represents a gain of $2/3 \times 4^n$ CNOTs and $2/3 \times 4^n$ rotations if we use rotation multiplexors, bringing the total to $2 \times 4^n$ CNOTs and $2 \times 4^n$ rotations. If we use multiplexors in $SU(2)$, only multiplexors on $k$ qubits are merging and not diagonal anymore, saving twice less CNOTs but the same number of rotations. We save in total $1/3 \times 4^n$ CNOTs and $2/3 \times 4^n$ rotations and the number of CNOTs, resp. rotations, becomes asymptotically equal to $4^n$, resp. $2 \times 4^n$. We also notice that the operators $X$ that appear when we switch to synthesis on a lower number of qubits disappear too by multiplying themselves. An example on $3$ qubits is showed in Figure \ref{householder-final-circuit}. We use the following notation: $|v_k|$ (resp. $|v_k^T|$) represents the operator that prepares (resp. deprepares) the real state $|v_k|$ consisting of the amplitudes of the components of the state $v_k$. $D_k$ is the diagonal gate containing the phases of the components of the state $v_k$ and $D_k^j = D_j^* \times D_k$. The results for the final gate counts are given Table \ref{gate_count}.

\subsubsection{Flop counts}
Apart from the circuit size, the other measure we are interested in is
the computational cost, measured in flops. 

The computational cost of the synthesis part is negligible
compared to the cost of the Householder decomposition. Overall state
preparations of states of size $2^n, 2^n-1, \hdots, 3, 2$ need to be
performed. For a state on $k$ qubits, it requires $O(k2^k)$
operations, and we need to do it for $2^{k-1}$ states. Thus the
synthesis part needs around
\[ \sum_{k=1}^n k2^k \times 2^{k-1} = O(n4^{n})  \] floating point
operations. This is asymptotically negligible compared to the
Householder factorization where $O(8^n)$ operations are needed.
Table~\ref{flop_count} summarizes the flop count for the various
methods.

\begin{table}
\centering
\begin{tabular}{lcc} \toprule
    {Method} & {flops} \\ \midrule
    Quantum Shannon & $19 \times 8^n$   \\[.9ex]
    Householder &  $2/3 \times 8^n$ \\ \midrule 
    Classical QR factorization & $4/3 \times 8^n$ \\\bottomrule
\end{tabular}
\caption{Asymptotic flop counts for decomposition methods}
\label{flop_count}
\end{table}

\section{Experimental results}\label{sec:exp}

The experiments have been carried out on one node of the QLM (Quantum Learning Machine) located at ATOS/BULL.
This node is a 24-core Intel Xeon(R) E7-8890 v4 processor at 2.4 GHz. Hyper-threading has been disabled.

Most of the programs are written in C with the C-interface for LAPACK~\cite{LAPACK} (LAPACKE). We adapted the LAPACK routine ZGEQRF (in Fortran) to compute the QR factorization of unitary matrices using the blocked algorithm described in Section~\ref{sec:householder}. LAPACK is linked with the MKL~\cite{MKL} multithreaded BLAS. The original ZGEQRF routine computes the best block size according to the size of the matrix and the hardware, we keep this computation in our modified routine. Our experiments use random unitary matrices generated via the LAPACK routine ZLAROR which generate matrices from a uniform distribution according to the Haar measure \cite{doi:10.1137/0717034}. This way we get the most generic matrices possible: dense, without any particular structure or pattern in the matrix elements. We are thus ensured to have a worst case scenario in terms of performance for our algorithms.

We present here numerical experiments to evaluate successively the sequential performance, the strong scalability and the weak scalability using multiple cores.

\subsection{Sequential runs}

In Figure~\ref{sequential} we compare the performance (in time) of the following routines or programs:
\begin{itemize}
  \item The LAPACK routine ZGEQRF that computes the QR factorization of a complex matrix in double precision (note that here the matrix is square), to serve as a reference.
  \item Our modified ZGEQRF routine adapted for unitary matrices.
  \item The complete circuit synthesis process which includes the QR factorization and the synthesis of the circuit obtained from this decomposition as explained in Section~\ref{sec:circuit}.
  \item The Quantum Shannon Decomposition (QSD), where the implementation essentially relies on the methodology described in~\cite{1629135} and uses the LAPACK routine ZUNCSD to compute the Cosine-Sine Decomposition (CSD). The routine implements the algorithm in \cite{Sutton2009}. This algorithm is the state of the art and has already been used in other implementations \cite{iten2019introduction,wang2013physical}.
\end{itemize}

\def\setplot#1#2#3{\node [#3] at #2 {#1};\def\lab{#1}\def\r{#2}\def\pc{#3}}
\def\plotdata#1{\node [\pc] at #1 {\lab}; \draw[very thick, dotted, color=\pc] \r -- #1; \def\r{#1}}

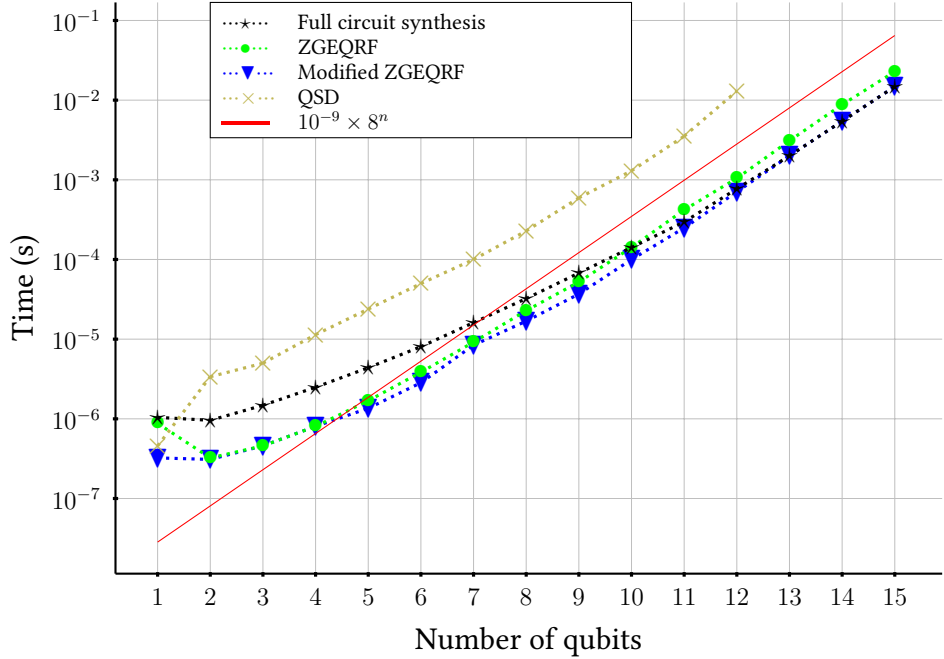
\begin{figure}[tbp]
  \centering
  \begin{tikzpicture}[x=7mm,y=.2mm]
    \draw[xstep=1,ystep=53,gray!50,very thin] (0.2,-50) grid (15.9,330);
    \draw[thick,-] (0.2,-50) -- node [anchor=north, inner sep=7mm] {Number of qubits} (15.9,-50);
    \draw[thick,-] (0.2,-50) -- node[anchor=east, inner sep=10mm] {\rotatebox{90}{Time (s)}} (0.2,330);
    \foreach \x in {1,2,3,4,5,6,7,8,9,10,11,12,13,14,15}
    \draw[very thick] (\x,-49) -- (\x,-52) node[anchor=north] {\scalebox{.8}{$\x$}};
    \foreach \x in {0,1,2,3,4,5,6}
    \pgfmathtruncatemacro\result{\x-7}
    \pgfmathtruncatemacro\pos{53*\x}
    \draw[very thick] (0.25,\pos) -- (0.15,\pos) node[anchor=east] {\scalebox{.8}{$10^{\result}$}};
    \def\p#1#2{\plotdata{(#1,381-#2)}}
    \setplot{$\blacktriangledown$}{(1,381-354)}{blue}\p{2}{355}\p{3}{346}\p{4}{333}\p{5}{321}\p{6}{304}\p{7}{279}\p{8}{263}\p{9}{245}\p{10}{222}\p{11}{201}\p{12}{177}\p{13}{153}\p{14}{130}\p{15}{107}
    \setplot{$\bullet$}{(1,381-331)}{green}\p{2}{354}\p{3}{346}\p{4}{333}\p{5}{316}\p{6}{297}\p{7}{277}\p{8}{256}\p{9}{237}\p{10}{214}\p{11}{189}\p{12}{168}\p{13}{143}\p{14}{119}\p{15}{97}
    \setplot{$\star$}{(1,381-327)}{black}\p{2}{329}\p{3}{319}\p{4}{307}\p{5}{294}\p{6}{280}\p{7}{264}\p{8}{248}\p{9}{231}\p{10}{214}\p{11}{197}\p{12}{175}\p{13}{153}\p{14}{130}\p{15}{107}
    \setplot{$\times$}{(1,381-346)}{yellow!70!black}\p{2}{300}\p{3}{291}\p{4}{272}\p{5}{255}\p{6}{238}\p{7}{222}\p{8}{203}\p{9}{181}\p{10}{163}\p{11}{140}\p{12}{110}

    \draw [red] (1,381-410)  -- (15,381-73);
    
    \draw[fill=white] (2,330) rectangle (10,240) ;
    \node at (2,332) [anchor=north west]
    {\scalebox{.7}{\begin{minipage}{6cm}
          \begin{tabular}{@{}ll@{}}
            \textcolor{black}{${\cdot}{\cdot}{\cdot}{\star}{\cdot}{\cdot}{\cdot}$}
            & Full circuit synthesis                              
            \\[-.2ex]
            \textcolor{green}{${\cdot}{\cdot}{\cdot}{\bullet}{\cdot}{\cdot}{\cdot}$}
            & ZGEQRF
            \\[-.2ex]
            \textcolor{blue}{${\cdot}{\cdot}{\cdot}{\blacktriangledown}{\cdot}{\cdot}{\cdot}$}
            & Modified ZGEQRF
            \\[-.2ex]
            \textcolor{yellow!70!black}{${\cdot}{\cdot}{\cdot}{\times}{\cdot}{\cdot}{\cdot}$}
            & QSD
            \\[-.2ex]
            \textcolor{red}{\raisebox{.5ex}{\rule{5ex}{1.5pt}}}
            & $10^{-9}\times 8^n$
            \\[-.2ex]
          \end{tabular}
        \end{minipage}}} ;
  \end{tikzpicture}
\caption{Sequential time for operator decomposition and circuit synthesis.}
\label{sequential}
\end{figure}

We considered matrices of sizes
$2^k \times 2^k, k=1 \dots15$ (operators acting on 1 to 15
qubits). The upper limit of 15 was chosen so that all decompositions can be achieved within an hour. This is why the curve plotting the QSD decomposition stops for 12 qubits.

As expected all the methods follow asymptotically $O(8^n)$ (curve also plotted) in accordance with the theoretical complexity.
The gap between the general and modified QR
factorizations in log scale
corresponds approximatively to a factor of 2, in accordance with the flop count.

When comparing the QR and QSD methods, we observe that for the same amount of time we can synthesize matrices with 2, almost 3 qubits more. The ratio between the times taken by the QSD and our method is even increasing with the number of qubits, reaching a value of almost 300 for $12$ qubits which is much bigger than the expected ratio of $30$. This is due to the routine ZUNCSD that does not follow the theoretical complexity and does not scale well with the number of qubits.

\subsection{Multithreaded runs}

\begin{figure}[tbp]
\centering
  \begin{tikzpicture}[x=4mm,y=.2mm]
    \draw[xstep=5,ystep=72,gray!50,very thin] (0,-20) grid (25,360);
    \draw[thick,-] (0,-20) -- node [anchor=north, inner sep=7mm] {Number of threads} (25,-20);
    \draw[thick,-] (0,-20) -- node[anchor=east, inner sep=10mm] {\rotatebox{90}{Time (s)}} (0,360);
    \draw[thick,-] (25,-20) -- node[anchor=west, inner sep=10mm] {\rotatebox{90}{Gflop/s}} (25,360);
    \foreach \x in {0,5,10,15,20,25}
    \draw[very thick] (\x,-18) -- (\x,-22) node[anchor=north] {\scalebox{.8}{$\x$}};
    \foreach \x in {0,1,2,3,4,5}
    \pgfmathtruncatemacro\result{\x*1000}
    \pgfmathtruncatemacro\pos{72*\x}
    \draw[very thick] (0.1,\pos) -- (-0.1,\pos) node[anchor=east] {\scalebox{.8}{${\result}$}};
    \foreach \x in {0,1,2,3,4,5,6,7,8}
    \pgfmathtruncatemacro\result{\x*25}
    \pgfmathtruncatemacro\pos{45*\x}
    \draw[very thick] (24.9,\pos) -- (25.1,\pos) node[anchor=west] {\scalebox{.8}{${\result}$}};

    \def\p#1#2{\plotdata{(#1,415-#2)}}
    \setplot{$\blacktriangledown$}{(1,415-401)}{blue}
    \p{2}{379}\p{4}{344}\p{6}{313}\p{8}{287}\p{10}{259}\p{12}{228}\p{14}{197}\p{16}{168}\p{18}{141}\p{20}{119}\p{22}{105}\p{24}{87}
    \setplot{$\blacktriangledown$}{(1,415-275)}{blue}
    \p{2}{345}\p{4}{379}\p{6}{390}\p{8}{396}\p{10}{400}\p{12}{403}\p{14}{405}\p{16}{407}\p{18}{409}\p{20}{409}\p{22}{409}\p{24}{409}
    \setplot{$\bullet$}{(1,415-406)}{green}
    \p{2}{393}\p{4}{363}\p{6}{334}\p{8}{310}\p{10}{287}\p{12}{262}\p{14}{242}\p{16}{220}\p{18}{202}\p{20}{187}\p{22}{176}\p{24}{163}
    \setplot{$\bullet$}{(1,415-63)}{green}
    \p{2}{220}\p{4}{316}\p{6}{352}\p{8}{366}\p{10}{375}\p{12}{381}\p{14}{386}\p{16}{389}\p{18}{392}\p{20}{394}\p{22}{395}\p{24}{396}
    \setplot{$\cdot$}{(1,415-401)}{red}
    \p{2}{379}\p{4}{344}\p{6}{310}\p{8}{282}\p{10}{254}\p{12}{219}\p{14}{187}\p{16}{157}\p{18}{126}\p{20}{105}\p{22}{84}\p{24}{69}
    \setplot{$\star$}{(1,415-251)}{black}
    \p{2}{329}\p{4}{371}\p{6}{383}\p{8}{389}\p{10}{392}\p{12}{395}\p{14}{397}\p{16}{399}\p{18}{400}\p{20}{401}\p{22}{402}\p{24}{402}
    
    \draw[fill=white] (2,330) rectangle (12,255) ;
    \node at (2,332) [anchor=north west]
    {\scalebox{.7}{\begin{minipage}{6cm}
          \begin{tabular}{@{}ll@{}}
            \textcolor{black}{${\cdot}{\cdot}{\cdot}{\star}{\cdot}{\cdot}{\cdot}$}
            & Full circuit synthesis                              
            \\[-.2ex]
            \textcolor{green}{${\cdot}{\cdot}{\cdot}{\bullet}{\cdot}{\cdot}{\cdot}$}
            & ZGEQRF
            \\[-.2ex]
            \textcolor{blue}{${\cdot}{\cdot}{\cdot}{\blacktriangledown}{\cdot}{\cdot}{\cdot}$}
            & Modified ZGEQRF
            \\[-.2ex]
            \textcolor{red}{${\cdot}{\cdot}{\cdot}{\cdot}{\cdot}{\cdot}{\cdot}{\cdot}$}
            & ZGEMM peak
          \end{tabular}
        \end{minipage}}} ;
  \end{tikzpicture}
\caption{Strong scaling for quantum operator decomposition and circuit synthesis on 15 qubits.}
\label{strong}
\end{figure}
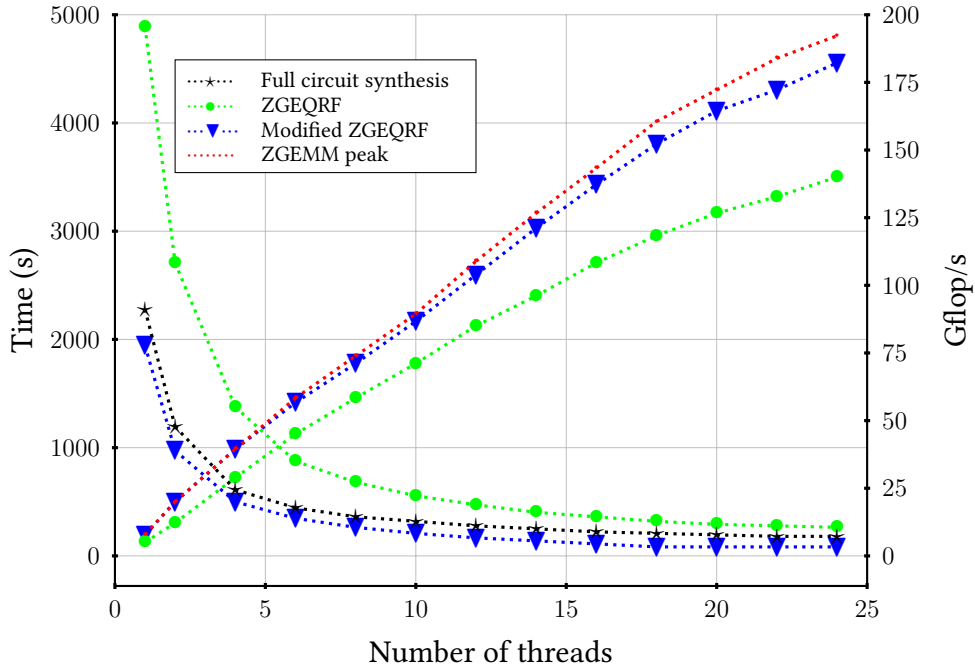

\begin{figure}[tbp]
  \centering
  \begin{tikzpicture}[x=4.5cm,y=.2mm]
    \draw[xstep=1,ystep=48,gray!50,very thin] (-.2,-40) grid (2.2,320);
    \draw[thick,-] (-.2,-40) -- node [anchor=north, inner sep=7mm]
    {Number of qubits $|$ Number of threads} (2.2,-40);
    \draw[thick,-] (-.2,-40) -- node[anchor=east, inner sep=10mm] {\rotatebox{90}{Gflop/s}} (-.2,320);
    \foreach \x in {0,1,2}
    \pgfmathtruncatemacro\y{2^(2*\x)}
    \pgfmathtruncatemacro\z{\x+13}
    \draw[very thick] (\x,-38) -- (\x,-42) node[anchor=north] {\scalebox{.8}{$\z~|~\y$}};
    \foreach \x in {0,1,2,3,4,5,6}
    \pgfmathtruncatemacro\result{(\x+1)*20}
    \pgfmathtruncatemacro\pos{48*\x}
    \draw[very thick] (-0.18,\pos) -- (-0.22,\pos) node[anchor=east] {\scalebox{.8}{${\result}$}};

    \setplot{$\blacktriangledown$}{(0,383-402)}{blue}\plotdata{(1,383-336)}\plotdata{(2,383-88)}
    \setplot{$\bullet$}{(0,383-410)}{green}\plotdata{(1,383-348)}\plotdata{(2,383-159)}
    \setplot{+}{(0,383-406)}{red}\plotdata{(1,383-326)}\plotdata{(2,383-73)}
    
    \draw[fill=white] (0.2,300) rectangle (1.2,235) ;
    \node at (0.2,302) [anchor=north west]
    {\scalebox{.7}{\begin{minipage}{6cm}
          \begin{tabular}{@{}ll@{}}
            \textcolor{blue}{${\cdot}{\cdot}{\cdot}{\blacktriangledown}{\cdot}{\cdot}{\cdot}$}
            & Modified ZGEQRF
            \\[-.2ex]
            \textcolor{green}{${\cdot}{\cdot}{\cdot}{\bullet}{\cdot}{\cdot}{\cdot}$}
            & ZGEQRF
            \\[-.2ex]
            \textcolor{red}{${\cdot}{\cdot}{\cdot}${+}${\cdot}{\cdot}{\cdot}{\cdot}$}
            & ZGEMM peak
          \end{tabular}
        \end{minipage}}} ;
  \end{tikzpicture}
  \caption{Weak scaling for quantum operator decomposition on 15 qubits.}
  \label{weak}
\end{figure}
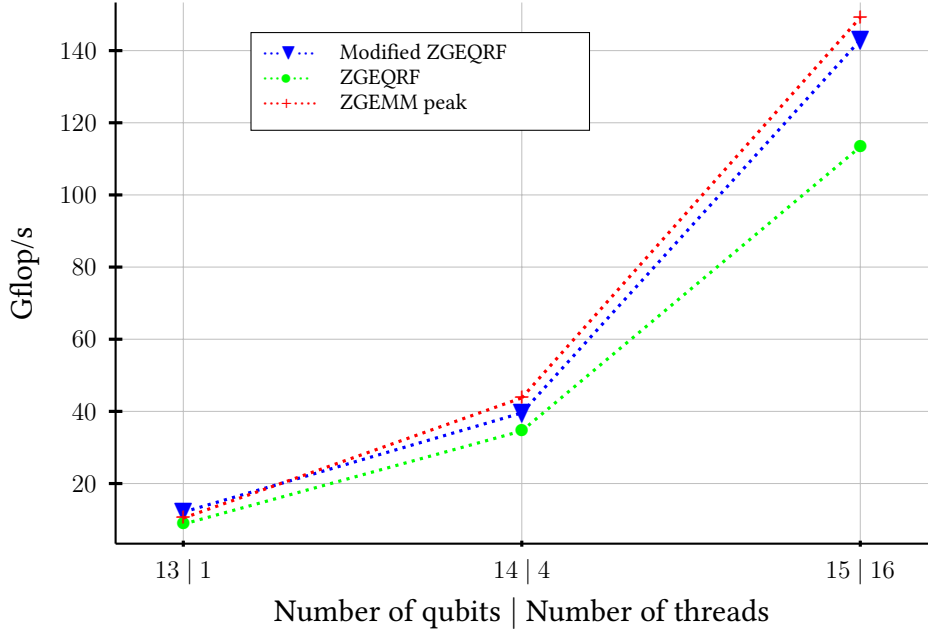

Because we could reach 15 qubits (unitary matrices of size $32768 \times 32768$) with a sequential run in less than one hour, we chose this size for our multithreaded runs.
The strong scalability is then evaluated using up to 24 threads.
Since the ZUNCSD routine used for the QSD is not parallel, it has been excluded from our experiments.
Figure~\ref{strong} presents performance results (in time and Gflop/s) for the chosen number of threads.
The time of the modified ZGEQRF scales like the full circuit synthesis since the QR factorization  represents most of the computational cost in the synthesis. Also, due to a smaller flop count, the modified QR is always much faster than the general QR. Moreover, looking at the Gflop/s performance rate, we observe that our modified QR factorization offers a good scalability due to an algorithm which is rich in BLAS 3 operations and provides a performance close to that of a matrix-matrix product (ZGEMM routine, also plotted in Figure~\ref{strong}).
Note that the Gflop/s rate for the full circuit synthesis is not plotted since the bulk of the arithmetical operations correspond to those of the factorization and the time of the synthesis itself is negligible.

Our experiments on weak scaling aim at measuring how the performance evolves with the number of threads but with a fixed problem size for each thread. Our algorithm for circuit synthesis can only accept matrices of size $2^n \times 2^n$ i.e. $4^n$ entries. As we can only multiply the size of our problems by a factor of $4$, we need to multiply also the number of threads by a factor of $4$. Thus, starting from a sequential run on $13$ qubits, we achieved experiments on $14$ and $15$ qubits using $4$ and $16$ threads respectively. The results given in Figure~\ref{weak} show that the rate (in Gflop/s) of the modified ZGEQRF increases with the number of threads/qubits with a very good scalability (close to that of ZGEMM) due to the mostly BLAS 3 operations implemented in the algorithm.

\subsection{Experiments on Graphics Processing Units (GPU)}

We performed additional experiments to study the behavior of our QR algorithm for unitary matrices using two Kepler K40 with 2880 CUDA cores and a multicore host composed of two Intel Xeon E5-2620 processors (6 cores each). The time for the synthesis is not plotted here since it is negligible compared to the time of the QR factorization.

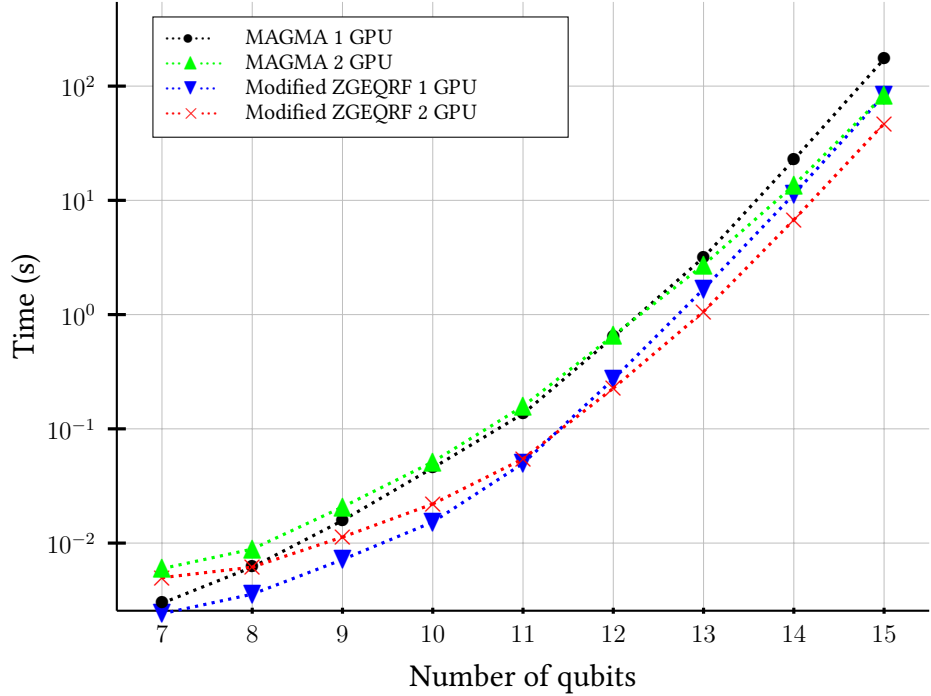
\begin{figure*}[t!]
    \centering
    \centering
    \begin{tikzpicture}[x=1.2cm,y=.2mm]
    \draw[xstep=1,ystep=76,gray!50,very thin] (6.5,-45) grid (15.5,360);
    \draw[thick,-] (6.5,-45) -- node[anchor=north, inner sep=7mm] {Number of qubits} (15.5,-45);
    \draw[thick,-] (6.5,-45) -- node[anchor=east, inner sep=10mm] {\rotatebox{90}{Time (s)}} (6.5,360);
    \foreach \x in {7,8,9,10,11,12,13,14,15}
    \draw[very thick] (\x,-43) -- (\x,-47) node[anchor=north] {\scalebox{.8}{$\x$}};
    \foreach \x in {0,1,2,3,4}
    \pgfmathtruncatemacro\result{\x-2}
    \pgfmathtruncatemacro\pos{\x*76}
    \draw[very thick] (6.6,\pos) -- (6.4,\pos) node[anchor=east] {\scalebox{.8}{$10^{\result}$}};
       
    \def\p#1#2{\plotdata{(#1,381-#2)}}
    \setplot{$\blacktriangledown$}{(7,381-428)}{blue}\p{8}{415}\p{9}{392}\p{10}{367}\p{11}{328}\p{12}{272}\p{13}{212}\p{14}{149}\p{15}{83}
    \setplot{$\bullet$}{(7,381-421)}{black}\p{8}{397}\p{9}{366}\p{10}{331}\p{11}{295}\p{12}{244}\p{13}{191}\p{14}{126}\p{15}{59}
    \setplot{$\times$}{(7,381-404)}{red}\p{8}{397}\p{9}{377}\p{10}{355}\p{11}{325}\p{12}{278}\p{13}{227}\p{14}{166}\p{15}{102}
    \setplot{$\blacktriangle$}{(7,381-398)}{green}\p{8}{385}\p{9}{357}\p{10}{327}\p{11}{290}\p{12}{243}\p{13}{196}\p{14}{143}\p{15}{83}

    \draw[fill=white] (6.9,350) rectangle (11.5,270) ;
    \node at (6.9,352) [anchor=north west]
    {\scalebox{.7}{\begin{minipage}{6cm}
          \begin{tabular}{@{}ll@{}}
            \textcolor{black}{${\cdot}{\cdot}{\cdot}{\bullet}{\cdot}{\cdot}{\cdot}$}
            & MAGMA 1 GPU
            \\[-.2ex]
            \textcolor{green}{${\cdot}{\cdot}{\cdot}{\blacktriangle}{\cdot}{\cdot}{\cdot}{\cdot}$}
            & MAGMA 2 GPU
            \\[-.2ex]
            \textcolor{blue}{${\cdot}{\cdot}{\cdot}{\blacktriangledown}{\cdot}{\cdot}{\cdot}$}
            & Modified ZGEQRF 1 GPU
            \\[-.2ex]
            \textcolor{red}{${\cdot}{\cdot}{\cdot}{\times}{\cdot}{\cdot}{\cdot}$}
            & Modified ZGEQRF 2 GPU
          \end{tabular}
        \end{minipage}}} ;
  \end{tikzpicture}
  \label{Kepler}
   \caption{Time for factorization of unitary matrices on GPUs.}
    \label{GPU_graph}
\end{figure*}

Similarly to what was made previously with the LAPACK routine, we modified the QR routine from the MAGMA~\cite{TDB.10} linear algebra library for GPUs according to Algorithm 3.1. Note that the transfer of the panel (block column factorized at each iteration) from the CPU to the GPU performed in MAGMA is replaced by a transfer of the 2 triangular matrices mentioned in Section~\ref{sec:householder} which are broadcasted to the GPUs involved in the computation.
In Figure~\ref{GPU_graph}, we obtain the factor of $2$ (due to twice less flops) between the standard and the modified QR factorization. We also observe that using 2 GPUs has no interest for problems smaller than 12 qubits but we get a factor close to 2 (e.g., 1.84 for 15 qubits) when switching from 1 to 2 GPUs for problems larger than 13 qubits, showing a good scalability of the algorithm.
 
\section{Conclusion}\label{sec:concl}

In this work we recalled the fundamentals of quantum computing and we stated the problem of quantum circuit synthesis. We highlighted the importance of having an efficient circuit synthesis framework by considering metrics based on flop and gate counts. To address this issue we presented a modified QR factorization in complex arithmetic based on Householder transformations where we  
exploit the specificities of unitary matrices to require twice as less flops and  
we proposed a scalable blocked implementation that contains mostly level-3 BLAS operations.
Then we described a method to convert the QR factorization into a quantum circuit with clearly defined properties. 
Our method results in a significant gain in time compared to the best methods in quantum compiling. 
As future work, we will study the behavior of our method on bigger problems using large distributed HPC systems. In terms of circuit size, some improvements may be obtained by studying in more details the optimization of state preparation occurring during the process.

\section*{Acknowledgement}
This work was supported in part by the French National Research Agency
(ANR) under the research project SoftQPRO ANR-17-CE25-0009-02,
and by the DGE of the French Ministry of Industry under the research
project PIA-GDN/QuantEx P163746-484124.

\end{document}